\documentclass[sn-basic]{sn-jnl}% Basic Springer Nature Reference Style/Chemistry Reference Style
\jyear{2021}%

\usepackage{natbib}
\usepackage{subfigure}
\usepackage{bm}
\usepackage{booktabs}

\def\argmin{\mathop{\rm argmin}}

\def\sup{\mathop{\rm sup}}

\providecommand{\norm}[1]{\lVert#1\rVert}

\def\ed{ \end{document} }

\newcommand{\ww}{39mm}

\newcommand{\hhhh}{35mm}
\newcommand{\wwww}{37mm}

\newdimen\Lwidth
\renewcommand\theenumi{\roman{enumi}}

\theoremstyle{thmstyleone}%
\theoremstyle{thmstyletwo}%

\theoremstyle{thmstylethree}%

\begin{document}

\title[Proximal nested sampling]{Proximal nested sampling for high-dimensional Bayesian model selection}


\author*[1,2]{\fnm{Xiaohao}~\sur{Cai}}\email{x.cai@soton.ac.uk}

\author*[1,3]{\fnm{Jason}~D.~\sur{McEwen}}\email{jason.mcewen@ucl.ac.uk}

\author*[4]{\fnm{Marcelo}~\sur{Pereyra}}\email{m.pereyra@hw.ac.uk}

\affil[1]{\orgdiv{Mullard Space Science Laboratory (MSSL)}, \orgname{University College London (UCL)}, \orgaddress{\city{Dorking}, \postcode{RH5 6NT}, \country{UK}}}
\affil[2]{\orgdiv{School of Electronics and Computer Science}, \orgname{University of Southampton}, \orgaddress{\city{Southampton}, \postcode{SO17 1BJ}, \country{UK}}}
\affil[3]{\orgname{Alan Turing Institute}, \orgaddress{\city{London}, \postcode{NW1 2DB}, \country{UK}}}
\affil[4]{\orgdiv{School of Mathematical and Computer Sciences}, \orgname{Heriot-Watt University}, \orgaddress{\city{Edinburgh}, \postcode{EH14 4AS}, \country{UK}}}


\abstract{Bayesian model selection provides a powerful framework for objectively comparing models directly from observed data, without reference to ground truth data. However, Bayesian model selection requires the computation of the marginal likelihood (model evidence), which is computationally challenging, prohibiting its use in many high-dimensional Bayesian inverse problems. With Bayesian imaging applications in mind, in this work we present the \textit{proximal nested sampling} methodology to objectively compare alternative Bayesian imaging models for applications that use images to inform decisions under uncertainty.  The methodology is based on nested sampling, a Monte Carlo approach specialised for model comparison, and exploits proximal Markov chain Monte Carlo techniques to scale efficiently to large problems and to tackle models that are log-concave and not necessarily smooth (e.g., involving $\ell_1$ or total-variation priors). The proposed approach can be applied computationally to problems of dimension $\mathcal{O}(10^6)$ and beyond, making it suitable for high-dimensional inverse imaging problems.  It is validated on large Gaussian models, for which the likelihood is available analytically, and subsequently illustrated on a range of imaging problems where it is used to analyse different choices of dictionary and measurement model.}

\keywords{Nested sampling, MCMC sampling, marginal likelihood, Bayesian evidence, inverse problems, proximal optimisation, model selection}



\maketitle

%-------------------------------------------------------------------
\section{Introduction}\label{sec:introduction}
%-------------------------------------------------------------------
High-dimensional inverse problems are ubiquitous in the data and imaging sciences, as well as in the physical and engineering sciences more generally. Due to limitations of the data observation process and measurement noise, or even just due to the nature of the problem at hand, most inverse problems encountered are seriously ill-conditioned or ill-posed (canonical examples include, e.g.,\ medical and radio interferometric imaging; \citealt{DMP16,CPraM17,ZYZL20,LHTSA21}). Developing better methodology for solving challenging inverse problems is a significant focus of the community. The Bayesian statistical framework is currently one of the predominant frameworks to perform inference in inverse problems \citep{RC04,PereyraIEEE16}. The choice of the Bayesian model used has a profound impact on the solutions delivered, as alternative models can lead to significantly different point estimations and uncertainty quantification results.

In this article we develop methodology to objectively compare alternative Bayesian models in performing inference in the regime of high-dimensional inverse problems, directly form the observed data and in the absence of ground truth. Motivated by applications in computational imaging, we focus on the comparison of models with posterior distributions that are log-concave and potentially not smooth. In this context, model selection has been traditionally addressed through benchmark experiments involving ground truth data and expert supervision. However, for many applications it is difficult and expensive to produce reliable ground truth data.  Moreover, for many problems it is simply impossible. Bayesian model selection provides a framework for selecting the most appropriate model directly from the observed data in an objective manner and without reference to ground truth data.

Bayesian model selection requires the computation of the \textit{marginal likelihood} of the data -- the average likelihood of a model over its prior probability space -- which is also called the \textit{Bayesian evidence}.  This quantity is a key ingredient of model selection statistics such as Bayes factors and likelihood ratio tests \citep{R01}. The computation of the marginal likelihood for high-dimensional models is highly non-trivial because it requires the computation of integrals over the (high-dimensional) solution space. For example, in the context of Bayesian imaging problems, the dimension is given by the number of parameters (e.g.\ pixels) of interest, which frequently reach sizes of $\mathcal{O}(10^5)$ to $\mathcal{O}(10^6)$ and beyond. For such settings, the evaluation of the marginal likelihood has been previously considered to be computationally intractable.

Broadly speaking, general purpose Monte Carlo methods can only handle model selection tasks for problems of dimension $\mathcal{O}(10)$ to $\mathcal{O}(10^2)$ (for reviews see \citealt{CBBF07,FW12,Llorente2020}). Nested sampling \citep{S06}, a state-of-the-art Monte Carlo strategy designed specifically for model selection, has enabled model selection for moderate dimensional problems of size $\mathcal{O}(10^2)$ to $\mathcal{O}(10^3)$ \citep{MPL06,FH08,FHB09,BPC11,FS13,HHL15}. To the best of our knowledge, model selection for larger problems is currently possible only for models with very specific structures (e.g., conditionally Gaussian models; \citealt{harro2020}).

%In this article we consider methodology for objectively comparing several alternative Bayesian models to perform inference in high-dimensional inverse problems, directly from the observed data and without ground truth available. We pay special attention to applications in scientific imaging such as radio-interferometric astronomical imaging. Accordingly, we focus on the automatic comparison of very high-dimensional posterior distributions that are log-concave, as this is main class of models used in Bayesian imaging problems. The underlying convex geometry of these models allows the use of state-of-the art proximal optimisation and specialised proximal MCMC methods to perform Bayesian computation efficiently, even in very large scale settings. We leverage recent developments in this area of computational statistics to develop a proximal nested sampling methodology specialised for this class of models, which have been traditionally too large for advanced Bayesian model comparison techniques.

In this work, we address the difficult computation of the marginal likelihood by proposing a new methodology that carefully integrates nested sampling \citep{S06} with proximal Markov chain Monte Carlo (MCMC) \citep{M15,DMP16}. This leads to a \textit{proximal nested sampling} methodology specialised for comparing high-dimensional posterior distributions that are log-concave but potentially not smooth. The proposed approach can be applied computationally to log-concave models of dimension $\mathcal{O}(10^6)$ and beyond, making it suitable for model comparison in Bayesian imaging problems. We demonstrate the approach with a range of scientific imaging applications.

%(MERGE THIS PARAGRAPH WITH THE PARAGRAPH BEFORE "The remainder of this article...") In this article we consider methodology for objectively comparing several alternative Bayesian models to perform inference in high-dimensional inverse problems, directly from the observed data and without ground trugh available. We pay special attention to applications in imaging sciences, particularly scientific imaging problems such as radio-astronomy. Accordingly, we focus on the automatic comparison of very high-dimensional posterior distributions that are log-concave, as this is main class of models used in Bayesian imaging problems. The underlying convex geometry of these models allows the use of state-of-the art proximal optimisation and specialised proximal MCMC methods to perform Bayesian computation efficiently, even in very large scale settings. We leverage recent developments in this area of computational statistics to develop a proximal nested sampling methodology specialised for this class of models, which have been traditionally too large for advanced Bayesian model comparison techniques.

The remainder of the article is organised as follows. In Section~\ref{sec:baye-infe} we recall the Bayesian model selection approach, highlight the associated computational challenges, and discuss proximal MCMC methodology for Bayesian computation for inverse problems with an underlying convex geometry. Section~\ref{sec:NS} recalls the standard nested sampling method. Our proposed proximal nested sampling framework is presented
in general form in Section~\ref{sec:method}. In Section~\ref{sec:pns-explicit} explicit forms of proximal nested sampling are presented for common forms of the likelihood and prior that arise in imaging sciences. Experimental results validating the proposed method and showcasing its use in scientific imaging applications are reported in Section~\ref{sec:numerics}. Finally, we conclude in Section~\ref{sec:conclusions}.

%-------------------------------------------------------------------
\section{Bayesian inference for high-dimensional inverse problems}\label{sec:baye-infe}

In this section we briefly recall the Bayesian decision-theoretic approach to model comparison,  introduce some elements of convex analysis which are essential for our method, and review proximal MCMC methods, which are an important component of the proximal nested sampling methodology proposed in Section 4. We conclude the section by briefly explaining the computational difficulties encountered in high-dimensional Bayesian model selection and why it is necessary to develop new methodology for this task. {Readers familiar with Bayesian model selection and with proximal MCMC methodology may prefer to skip this section and continue reading from Section \ref{sec:NS}.}

\subsection{Bayesian estimation and model selection}
%-------------------------------------------------------------------
Let $\Omega \subseteq \mathbb{R}^d$. We consider the estimation of a quantity of interest $x \in \Omega$ from observed data $y$. Bayesian methods address such problems by postulating a statistical model ${\cal M}$ relating $x$ and $y$, from which estimators of $x$ and other inferences can be derived. More precisely, ${\cal M}$ defines a joint probability distribution $p(x, y \vert {\cal M})$ specified via the decomposition $p(x, y \vert {\cal M}) =  p(y \vert x, {\cal M}) p(x\vert{\cal M})$, where $p(y \vert x, {\cal M})$ denotes the likelihood of $x$ for the observed data $y$, and the marginal $p(x\vert{\cal M})$ is the so-called prior of $x$. Following Bayes' theorem, inferences on $x\vert y$ are then based on the posterior distribution
\begin{equation} \label{eqn:baye-thm}
	{p}(x \vert y, {\cal M}) =  \frac{{p}(y \vert x, {\cal M}) {p}(x \vert {\cal M})}{{p}(y \vert {\cal M})},
\end{equation}
which models our beliefs about $x$ after observing $y$. With applications in Bayesian imaging sciences in mind, we focus on posterior distributions that are log-concave and assume that the potential function $x \mapsto -\log {p}(x \vert y, {\cal M})$ is convex lower semicontinuous (l.s.c.) on $\Omega$, but possibly not smooth. This is an important class of models in modern Bayesian imaging sciences because it leads to point estimators that are by construction well-posed and that can be efficiently estimated by using scalable proximal convex optimisation and stochastic sampling methods \citep{KS05,RC04,PereyraIEEE16}.

We condition on ${\cal M}$ explicitly in \eqref{eqn:baye-thm} because our focus is model selection, where one entertains several alternative posterior distributions for $x \vert y$ stemming from different underlying modelling assumptions. As a result, rather than the posterior ${p}(x \vert y, {\cal M})$, our main object of interest is the {\it marginal likelihood} or \emph{model evidence}
\begin{equation} \label{eqn:marginal_likelihood}
	{p}(y \vert {\cal M}) = \int_{\Omega} p(y,x \vert {\cal M}) \text{d}x = \int_{\Omega} p(y \vert x, {\cal M}) p(x \vert {\cal M}) \text{d}x\, ,
\end{equation}
which measures the likelihood of the observed data under model ${\cal M}$, and which we use to objectively compare different models relating $x$ and $y$ \citep{R01}. Notice that the likelihood of the observed data $y$ under the model ${\cal M}$ is essentially the expectation (or average value) of the likelihood function $p(y \vert x,{\cal M})$ with respect to (w.r.t.) the prior $p(x \vert {\cal M})$. Therefore, a model that allocates its prior mass to solutions that agree with the observed data achieves a large marginal likelihood value. Conversely, a low marginal likelihood value indicates that only a small proportion of the solutions favoured by the prior agree with the observed data. In other words, the marginal likelihood \eqref{eqn:marginal_likelihood} measures the degree to which the observed data is in agreement with the assumptions of the model, and in doing so it provides a goodness-of-fit summary. Moreover, because all priors have the same total probability mass (i.e., $\int_\Omega p(x) \text{d}x = 1$), the likelihood \eqref{eqn:marginal_likelihood} naturally incorporates Occams's razor, trading off model simplicity and accuracy and penalising over-fitting \citep{R01}.

Bayesian model selection arises from the common and natural inquiry of which model is the most suitable to analyse $x \vert y$ from a set of models ${\cal M}_1,\ldots,{\cal M}_K$ available. For simplicity and without loss of generality, we suppose two alternative models ${\cal M}_1$ and ${\cal M}_2$ (the generalisation to additional models is straightforward). From Bayesian decision theory, to objectively compare the two models in settings without ground truth available, one should calculate the \textit{Bayes factor} \citep{R01}
\begin{equation} \label{eqn:post-ratio-complete}
	\rho_{12} =  \frac{{p}({\cal M}_1 \vert y)}{{p}({\cal M}_2 \vert y)}\frac{{p}({\cal M}_2)}{{p}({\cal M}_1)}
\end{equation}
where ${p}({\cal M}_1)$ and ${p}({\cal M}_2)$ denote the prior probabilities assigned to the two competing models, and where, from Bayes' theorem, we have that for any $i \in \{1,2\}$
\begin{equation}
	{p}({\cal M}_i \vert y) = \frac{p(y \vert {\cal M}_i) {p}({\cal M}_i)}{p(y \vert {\cal M}_1) {p}({\cal M}_1) + p(y \vert {\cal M}_2) {p}({\cal M}_2)}\,.
\end{equation}
By developing \eqref{eqn:post-ratio-complete} we can easily express the Bayes factor as the likelihood ratio
\begin{equation} \label{eqn:post-ratio-simple}
	\rho_{12} =  \frac{{p}({y\vert \cal M}_1)}{{p}(y \vert {\cal M}_2)},
\end{equation}
highlighting that $\rho_{12}$ is invariant to choice of the prior probabilities ${p}({\cal M}_1)$ and ${p}({\cal M}_2)$. If one assumes ${p}({\cal M}_1)={p}({\cal M}_2) = 1/2$ to reflect the absence of prior information, then the factor also coincides with the posterior probability ratio $p({\cal M}_1 \vert y)/p({\cal M}_2 \vert y)$.

Being a likelihood ratio, the factor $\rho_{12}$ is straightforward to read: if $\rho_{12} \gg 1$, we prefer model ${\cal M}_1$ over the alternative ${\cal M}_2$; conversely, if $\rho_{12} \ll 1$,
we prefer model ${\cal M}_2$; and if $\rho_{12} \approx 1$, we do not prefer either, inasmuch as the data $y$ are insufficient for us to make an informed judgement. The fact that $\rho_{12}$ is a likelihood ratio is also appealing from a frequentist viewpoint, as it is associated with the most powerful test for these two model hypotheses \citep{RC02}.

Unfortunately, calculating $\rho_{12}$ is generally not possible in large-scale settings because the dimensionality of $x$ renders the marginal likelihoods ${p}(y \vert {\cal M}_1)$ and ${p}(y  \vert {\cal M}_2)$ computationally intractable. More precisely, the marginal likelihoods are doubly-intractable because they require computing two intractable integrals over the space of solutions $\Omega$: the marginalisation of $x$ denoted explicitly in \eqref{eqn:marginal_likelihood}; and the normalising constant of the priors $p(x  \vert \mathcal{M}_i)$ when these are not available analytically, which otherwise implicitly also requires integrating over $\Omega$.

It is worth emphasising at this point that this major difficulty related to model selection is not encountered when performing inferences with the posteriors $p(x \vert y,\mathcal{M}_1)$ and $p(x \vert  y,\mathcal{M}_2)$ individually, as one can use MCMC methods to sample from $p(x \vert y,\mathcal{M})$ without ever having to evaluate the marginal likelihood $p(y\vert \mathcal{M})$. As a result, efficient Bayesian model selection remains an open problem in many areas of science and engineering that have widerly adopted Bayesian inference techniques for point estimation and uncertainty quantification.% This article focuses on the development of efficient Bayesian computation methodology for model selection in large-scale inverse problems with an underlying convex geometry. 

In the following we briefly recall MCMC sampling methods derived from the overdamped Langevin diffusion process, particularly proximal MCMC techniques specialised for large models that are log-concave, and explain why it is necesary to modify them to enable efficient model comparison.

%%%%%%%%%
\subsection{Bayesian computation and proximal MCMC methods}
\subsubsection{Convex analysis}
Let {$f : \mathbb{R}^d \rightarrow
		\left[-\infty , +\infty \right]$}. The function
	{$f$ is said to be proper if there exists $x_0 \in \mathbb{R}^d$ such that $f(x_0) < +\infty$.}
Denote for all $M \in \mathbb{R}$, $\{ f \leq M \} = \{ z \in
	\mathbb{R}^d \ \vert \ f(z) \leq M \}$. The function $f$ is  l.s.c. if for all $M \in \mathbb{R}$, $\{ f \leq M \}$ is a closed subset of
$\mathbb{R}^d$.
For $k \geq 0$, denoted by ${\cal C}^k(\mathbb{R}^d)$ the set of
$k$-times continuously differentiable functions. For $f \in
	{\cal C}^1(\mathbb{R}^d)$, denote by $\nabla f$ the gradient of $f$. We say that $f \in {\cal C}^1(\mathbb{R}^d)$ is a Lipschitz continuously differentiable function if there exists $C \geq 0$ such that for all $x,y \in \mathbb{R}^d$, $\norm{\nabla f(x) - \nabla f(y)} \leq C \norm{x-y}$.

Given a convex, proper, l.s.c. function $h: \mathbb{R}^d \rightarrow (-\infty, +\infty]$ and $\lambda > 0$,
the proximal operator \citep{BC11} associated with function $h$ at $x \in \mathbb{R}^d$ is defined as
\begin{equation} \label{eqn:prox-opt}
	\text{prox}^{\lambda}_h(x) = \argmin_{u\in \mathbb{R}^d} \big\{h(u) + \|u - x\|_2^2/2\lambda \big\}.
\end{equation}
When $\lambda=1$, we denote $\text{prox}^{1}_h(x)$ by $\text{prox}_h(x)$ for simplicity.

Let ${\cal K}$ be a closed convex set in $\mathbb{R}^d$ and let $\chi_{\cal K}$ be the characteristic function for $\cal K$, defined by $\chi_{\cal K}(x) = 0$ if $x \in \cal K$ and $+\infty$ otherwise. The proximal operator of $\chi_{\cal K}$ is the projection onto ${\cal K}$, given by
\begin{equation} \label{eqn:proj-opt}
	\text{proj}_{{\cal K}}(x) = \argmin_{u\in \mathbb{R}^d} \big\{\chi_{\cal K}(u) + \|u - x\|_2^2/2 \big\}\, .
\end{equation}

The convex conjugate of function $h$, denoted by $h^*$, is defined as
\begin{equation} \label{eqn:con-conj}
	h^*(x) = \sup_{u\in \mathbb{R}^d} \big\{x^\top u - h(u) \big\}.
\end{equation}
Its proximal operator can be related to the proximal operator of $h$ by
\begin{equation} \label{eqn:con-conj-prox}
	\text{prox}_{h^*}(x) = x - \text{prox}_{h}(x).
\end{equation}

The $\lambda$-Moreau-Yosida envelope of $h$ \citep{BC11} is given for any $x \in \mathbb{R}^d$ and $\lambda > 0$ by
\begin{equation} \label{eqn:prox-my-enve}
	h^{\lambda}(x) = \min_{u\in \mathbb{R}^d} \big\{h(u) + \|u - x\|_2^2/2\lambda \big\} .
\end{equation}
The envelope $h^{\lambda}$ is continuously differentiable with Lipschitz gradient. In particular, using the proximal operator, the gradient of $h^{\lambda}$ can be written
\begin{equation} \label{eqn:prox-diff}
	\nabla h^{\lambda}(x) = \big(x - \text{prox}^{\lambda}_h(x)\big) / \lambda,
\end{equation}
with $\lambda$ simultaneously controlling the Lipschitz constant of $\nabla h^{\lambda}$ as well as the error between $h$ and its smooth approximation $h^{\lambda}$. This approximation error can be made arbitrarily small by reducing $\lambda$, at the expense of deteriorating the regularity of $\nabla h^{\lambda}$, and consequently the speed of convergence of proximal Bayesian computation algorithms rely on $h^{\lambda}$.

%%%%%%%%
\subsubsection{Proximal Langevin MCMC sampling}
Consider the problem of calculating probabilities or expectations with respect to (w.r.t.) some distribution $\pi(\text{d}x)$ which admits a density $\pi(x)$ w.r.t.\  the usual $d$-dimensional Lebesgue measure. In the context of Bayesian inference, this is typically the posterior $p(x \vert y,\mathcal{M})$. Evaluating expectations and probabilities w.r.t.\  $\pi$ is non-trivial in problems of moderate and high dimension because of the integrals involved, which are usually beyond the scope of analytical and deterministic numerical integration schemes. These calculations are further complicated when the normalising constant of $\pi$ is not known, as this requires evaluating an additional $d$-dimensional integral. Monte Carlo sampling methods address these difficulties by simulating a set of samples from $\pi$ followed by Monte Carlo stochastic integration to compute probabilities and expectations w.r.t.\  $\pi$. While there are different ways of simulating samples from $\pi$, we focus on MCMC strategies where one proceeds by constructing a Markov chain that has $\pi$ as its invariant stationary distribution. Again, there are different methods for constructing such Markov chains (see \citealt{RC04} for an excellent introduction to MCMC methodology and \citealt{Green2015} for a survey of recent developments in the Bayesian computation literature).
% but most applications in inverse problems use the Metropolis-Hastings (MH) method \cite{RC04}, Gibbs sampling \cite{RC04}, Hamiltonian Monte Carlo \cite{N12}, and slice sampling \cite{N03}

The fastest provably convergent MCMC methods for Bayesian inference models can be derived from the Langevin diffusion process, which we recall below. For simplicity, rather than presenting the approach in full generality, we focus our presentation on proximal overdamped Langevin sampling for non-smooth models, which we later use in the proximal nested sampling method proposed in Section 4. For a more exhaustive introduction to the topic please see \citet[Section 2]{vargas2019accelerating} and references therein.

Assume that $\pi$ admits a decomposition $\pi(x) \propto \exp\{-f(x) - g(x)\}$ for all $x \in \mathbb{R}^d$, where $f \in \mathcal{C}^1(\mathbb{R}^d)$ with $\nabla f$ Lipschitz continuous with constant $L_f$, and where $g$ is a proper l.s.c. function that is convex on $\mathbb{R}^d$ but potentially non-smooth (e.g., $g$ could encode constraints on the solution space and involve non-smooth regularisers such as the $\ell_1$ norm). To simulate from $\pi$, we construct the overdamped Langevin stochastic differential equation (SDE) on $\mathbb{R}^d$ given by \citet{DMP16}
\begin{equation}\label{eqn:LangevinSDE}
	\text{d}X_t = -[\nabla f(\text{X}_t) +\nabla g^\lambda(\text{X}_t)] \text{d}t + \sqrt{2} \text{d}W_t \,, \quad X_0 = x_0\, ,
\end{equation}
where $(W_t)_{t\geq 0}$ is a $d$-dimensional Brownian motion, $g^\lambda$ is the Moreau-Yosida envelop of $g$ given by \eqref{eqn:prox-my-enve}, $\lambda > 0$ is a smoothing parameter that we will discuss later, and $x_0 \in \mathbb{R}^d$. {When $x \rightarrow f(x)+g^\lambda(x)$ is convex, the SDE has a unique strong solution and $X_t$ converges exponentially fast (as $t \rightarrow \infty$) to an invariant measure that is in the neighbourhood of $\pi$.}

To use \eqref{eqn:LangevinSDE} for Bayesian computation, we use a numerical solver to compute a discrete-time approximation of $X_t$ over some time period $t \in [0,T]$; the resulting discrete sample path constitutes our set of Monte Carlo samples. In particular, in this article we use the conventional Euler-Maruyama approximation
\begin{equation}\label{eqn:eulerMaruyDisc}
	X_{n+1} = X_{n} - \frac{\delta}{2}\nabla f(X_{n}) - \frac{\delta}{2}\nabla g^\lambda(X_{n}) + \sqrt{\delta} Z_{n+1},
\end{equation}
where $\delta \in [0,1/(L_f + 1/\lambda)]$ is a given stepsize and $(Z_{n})_{n\geq 1}$ is a sequence of i.i.d. \textit{d}-dimensional standard Gaussian random variables. This MCMC method is known as the Moreau-Yosida unadjusted Langevin algorithm (MYULA) \citep{DMP16}. The Markov chain \eqref{eqn:eulerMaruyDisc} is usually implemented by using \eqref{eqn:prox-diff} and reads
\begin{equation}\label{eqn:MYULADisc}
	X_{n+1} = X_{n} - \frac{\delta}{2} \nabla f (X_n)-\frac{\delta}{2\lambda}\left(X_{n}-\text{prox}_{g}^{\lambda}(X_{n})\right)+ \sqrt{\delta} Z_{n+1} \,.
\end{equation}
The smoothing parameter $\lambda$ and the stepsize $\delta$ jointly control a bias-variance trade-off between the asymptotic estimation errors and non-asymptotic errors associated with using a finite number of iterations. %Large values of $\lambda$ and $\delta$ lead to faster convergence to stationarity but also more asymptotic bias. 
In this article, we use $\lambda = 1/L_f$ and $\delta = 0.8/(L_f + 1/\lambda)$ as recommended in \citet{DMP16} (recall that $\nabla f$ is Lipschitz continuous with constant $L_f$, please see \citealt{DMP16,vargas2019accelerating} for further details).

The samples generated by \eqref{eqn:MYULADisc} can be directly used for biased Monte Carlo estimation \citep{DMP16}. Alternatively, at the expense of additional computation, one can supplement each iteration of MYULA with an MH (Metropolis-Hastings) correction step to asymptotically remove the approximation errors related to the discretisation of the SDE and the use of $g^\lambda$ instead of $g$, leading to a type of Metropolis-adjusted Langevin algorithm (MALA) (see \citealt{M15} for details).

%%%%%%%
\subsection{Estimation of marginal likelihoods and Bayes factors}
Let $\{X_{n}\}_{n=1}^N$ be a set of samples from $\pi$ (or an approximation of $\pi$), generated by using a proximal MCMC method or otherwise. Following a Monte Carlo integration approach, the expectation of any function $\phi : \mathbb{R}^d \rightarrow \mathbb{R}$ w.r.t.\  $\pi$ is approximated by
\begin{equation}\label{eqn:MC_estimator}
	\hat{\text{E}}_\pi (\phi) = \frac{1}{N}\sum_{n=1}^N \phi(X_{n})\, ,
\end{equation}
which, under assumptions, converges to the truth $\text{E}_\pi(\phi) = \int_\Omega \phi(x)\pi(x)\text{d}x$ as $N$ increases (or to a biased estimate if the samples are not exactly from $\pi$). The accuracy of Monte Carlo estimates depends of course on the number of samples $N$ and on the properties of the MCMC method used, but it also depends crucially on the variance $\text{Var}_\pi (\phi)$. Unfortunately, $\text{Var}_\pi (\phi)$ is often very large for the kinds of functions $\phi$ required for estimating the marginal likelihood \eqref{eqn:marginal_likelihood} (and in some cases $\text{Var}_\pi (\phi)$ is not even defined), leading to Monte Carlo estimators of the marginal that behave poorly \citep{NR94}. As a result, it is difficult to use the samples $\{X_{n}\}_{n=1}^N$ to perform model selection.
Several strategies have been proposed to address the aforementioned difficulty and derive well-posed estimators for the marginal likelihood \eqref{eqn:marginal_likelihood} and the Bayes factor \eqref{eqn:post-ratio-complete} (for reviews of classical methods see \citealt{FW12,CBBF07}).

One avenue is to generate samples from a sequence of distributions bridging $\pi$ to some tractable reference $\pi_0$ such as the prior distribution or a Gaussian approximation of $\pi$, e.g., thermodynamic integration \citep{RF96} and annealed important sampling \citep{N01}. Such strategies struggle with large problems because the number of intermediate distributions grows quickly as $d$ increases.

Another promising approach to derive computationally efficient estimators is to construct Rao-Blackwellized estimators by carefully introducing auxiliary variables, as proposed in the seminal papers \citet{C95} and \citet{CJ01}. This strategy has been successfully applied recently to signal and image processing models that are conditionally Gaussian given conjugate model hyper-parameters \citep{harro2020}. Some generalisations are possible, but constructing efficient Rao-Blackwellized estimators for more general classes of models, e.g., of the form \eqref{eqn:baye-thm}, is highly non-trivial.

An alternative natural strategy for stable Monte Carlo estimators for \eqref{eqn:marginal_likelihood} and \eqref{eqn:post-ratio-complete} is to construct a truncated estimator by first using the samples $\{X_{n}\}_{n=1}^N$ to identify a suitable truncating set $\mathcal{A}$, followed by a sample average \eqref{eqn:MC_estimator} only with the samples verifying $X_{n} \in \mathcal{A}$ \citep{BrosseCOLT2017}. Although by construction well-posed, truncated estimators need to be de-biased by using the volume of $\mathcal{A}$, which is usually very expensive to compute when the dimension $d$ is large. From the results of \citet{BrosseCOLT2017}, we believe that this strategy is unlikely to produce scalable methods suitable for large problems. One can circumvent or simplify the calculation of the volume of $\mathcal{A}$ (e.g., see \citealt{DMP16}), but in our experience the resulting estimators become unstable and are difficult to use.

Another alternative approach, which is agnostic to the sampling method, is the harmonic mean estimator \citep{NR94}; although, in its original form the variance of the estimator can be very poorly behaved such that the estimator can be highly inaccurate in practice.  Strategies to resolve this issue have been developed in the recently proposed learnt harmonic mean estimator \citep{mce20}, which has been shown to be highly effective and can scale to dimension $\mathcal{O}(10^3)$ and beyond.  Nevertheless, it may be challenging to scale this approach to the high-dimensional settings considered in this paper.

One can also consider the widely used Laplace's method \citep{TK86}, which relies on the assumption that the posterior distribution can be adequately approximated by a Gaussian distribution. Unfortunately, this is a strong assumption that often leads to inaccurate estimates in inverse problems that are ill-conditioned or ill-posed, particularly if $d \geq \textrm{dim}(y)$.
Many other alternatives are described in the literature, e.g., the Savage–Dickey density ratio \citep{T07} and Reversible Jump MCMC \citep{Green95}, which are mainly useful for nested or small models.
It is worth mentioning that there are also some model selection strategies that do not rely on the computation of the marginal likelihood (see, e.g., \citealt{kamary2018testing,PereyraEUSIPCO16}); however these are usually very computationally intensive.

Finally, nested sampling provides a distinctively different approach for efficiently estimating \eqref{eqn:marginal_likelihood} and \eqref{eqn:post-ratio-complete} \citep{S06}. The key idea underpinning nested sampling is the re-parameterisation of the marginal likelihood \eqref{eqn:marginal_likelihood} as a one-dimensional integral of the likelihood with respected to the enclosed prior volume.  This greatly reduces the computation costs involved, provided that one can efficiently sample from the prior distribution subject to a hard constraint on the likelihood value.
Nested sampling therefore shifts the computational challenge from the direct evaluation of a high-dimensional integral to sampling of the prior subject to a hard likelihood constraint.
The generation of samples is challenging and previous works have considered a range of sampling strategies. For example, conventional MCMC sampling \citep{S06}, rejection sampling (e.g.\ \citealt{MPL06,FH08,FHB09}), slice sampling (e.g.\ \citealt{HHL15}), and more advanced MCMC samplers such as Galilean Monte Carlo \citep{FS13} and diffusive nested sampling \citep{BPC11}.  Following over a decade of active research, nested sampling is now a well-established technique for computing the marginal likelihood that has found widespread application, particularly in astronomy \citep[e.g.][]{FH08,FHB09,T07}. Nevertheless, broadly speaking, current nested sampling techniques remain restricted to moderate dimensional problems of size $\mathcal{O}(10^2)$ to $\mathcal{O}(10^3)$.

With imaging problems in mind, this article presents an efficient nested sampling methodology specifically designed for high-dimensional log-concave models of the form \eqref{eqn:baye-thm}. A significant novelty of the proposed approach is that we address the difficult generation of samples by using a proximal MCMC technique that is naturally suited for dealing with high-dimensional log-concave distributions subject to hard convex constraints. Moreover, the proximal nature of the method straightforwardly allows the use of the non-smooth priors that are frequently encountered in imaging (e.g., involving the $\ell_1$ and total-variation regularisers), which would not be easily addressed by using alternative gradient-based samplers. Section~\ref{sec:NS} below reviews the nested sampling approach. The proposed proximal nested sampling methodology is presented in Section~\ref{sec:method}.

%-------------------------------------------------------------------
\section{Nested sampling}\label{sec:NS}
For ease of notation, given a model $\mathcal{M}$, let ${\cal L}(x) = p(y\vert x,\mathcal{M})$ denote the likelihood function, $\pi(x) = p(x\vert  \mathcal{M})$ the prior, and
\begin{align} \label{eqn:baye-evidende}
	\begin{split}
		{\cal Z} = p(y\vert \mathcal{M}) = \int_{\Omega} {\cal L}(x) \pi(x) \text{d} x,
	\end{split}
\end{align}
the marginal likelihood or evidence associated with a given model $\mathcal{M}$ (to simplify notation, we henceforth omit the dependence of ${\cal Z}$ and ${\cal L}$ on $y$).

Nested sampling \citep{S06} was proposed specifically to facilitate the efficient evaluation of ${\cal Z}$ for Bayesian model selection, while also supporting posterior inferences. As mentioned previously, the calculation of the multidimensional marginal likelihood integral \eqref{eqn:baye-evidende} is generally computationally intractable. Nested sampling addresses this difficulty by cleverly converting \eqref{eqn:baye-evidende} to a one-dimensional integral by re-parameterising the likelihood in terms of the enclosed prior volume. In addition, nested sampling involves the prior via simulation and hence does not require knowledge of the prior normalising constant. As a result, it also circumvents the second level of intractability of ${\cal Z}$ that arises in imaging problems.

Let $\Omega_{L^*} = \{x \vert {\cal L} (x) > L^* \}$, which groups the parameter space $\Omega$ into a series of nested subspaces
according to the level-set or iso-likelihood contour ${\cal L}(x) = L^* \ge 0$. Note that $\Omega_{L^* = 0} = \Omega$, since the likelihood
values cannot be negative. Define the prior volume $\xi$ by
%$\text{d} \xi = \pi(x) \text{d} x$, where
\begin{equation} \label{eqn:prior-vol}
	\xi(L^*) =  \int_{\Omega_{L^*}} \pi(x) \text{d} x.
\end{equation}
Note that $\xi(0) = 1$ and $\xi(L_\text{max}) = 0$, where $L_\text{max}$ is the maximum of the likelihood in $\Omega$.
Let {${\cal L}^\dagger(\xi)$} be the inverse of the prior volume $\xi(L^*)$ such that ${\cal L}^\dagger(\xi(L^*)) = L^*$\footnote{{In other words, ${{\cal L}^\dagger}$ is a tail quantile function such that, for any $L^* > 0$, the inverse of ${{\cal L}^\dagger}(\xi(L^*))$ represents the probability that a draw $x$ from the prior $\pi$ will have a likelihood ${\cal L}(x) > L^*$.}},
and assume it is a monotonically decreasing
function of $\xi$ (which, {when $\mathcal{L}$ is continuous and $\pi$ has connected support}, is satisfied theoretically and up to practical numerical considerations that can be trivially overcome; \citealt{SS06}).
The marginal likelihood integral \eqref{eqn:baye-evidende} can then be rewritten as
\begin{align} \label{eqn:baye-evidende-re}
	{\cal Z} & = \int_0^1 {\cal L}^\dagger(\xi) \text{d} \xi,
\end{align}
which is a one-dimensional integral over the prior volume $\xi$.

To evaluate \eqref{eqn:baye-evidende-re} in practice it is necessary to compute likelihood level-sets (iso-contours) $L_i$, which correspond to prior volumes $0< \xi_i \leq 1$ satisfying \eqref{eqn:prior-vol}.  A strategy to generate the likelihoods $L_i$ and associated prior volumes $\xi_i$ is discussed in Section~\ref{sec:NS-evidence-eval}.  Once the likelihoods $L_i = {\cal L}^\dagger(\xi_i)$ are obtained,  \eqref{eqn:baye-evidende-re} can be used to evaluate
the marginal likelihood, where $\{\xi_i\}_{i = 0}^{N}$ is a sequence of decreasing prior volumes, i.e.,
\begin{equation}
	0 < \xi_{N} < \cdots < \xi_1 < \xi_0 = 1.
\end{equation}
After discretising the integral \eqref{eqn:baye-evidende-re} and associating each likelihood $L_i$ a quadrature weight $w_i$,
the marginal likelihood can be computed
numerically using standard quadrature methods to give
\begin{equation} \label{eqn:baye-evi-compt}
	{\cal Z} \approx  \sum_{i=1}^{N} L_i w_i.
\end{equation}
The simplest assignment of the quadrature weights is $w_i = \xi_{i-1} - \xi_i$. The trapezium rule can also be used, i.e., $w_i = (\xi_{i-1} + \xi_{i+1})/2$. The approximation error related to the discretisation of \eqref{eqn:baye-evidende-re} can be made arbitrarily small by increasing $N$.

%-------------------------------------------
\subsection{Posterior inferences}\label{sec:NS-post-inf}
%-------------------------------------------
Posterior inferences can be easily computed once ${\cal Z}$ is
found. Any sample taken randomly in the prior volume interval $(\xi_{i-1}, \xi_i)$ is simply assigned an importance weight
\begin{equation} \label{eqn:post-inf}
	p_i =  \frac{L_i w_i}{\cal Z}.
\end{equation}
Samples with the assigned weights $\{p_i\}$ can then be used to calculate posterior inferences
such as the posterior moments, probabilities, and credible regions.

%-------------------------------------------
\subsection{Marginal likelihood evaluation}\label{sec:NS-evidence-eval}
%-------------------------------------------
We now recall the basic procedure of the standard nested sampling framework for evaluating the marginal likelihood, i.e.\ to compute the summation \eqref{eqn:baye-evi-compt}.  In particular, it is necessary to generate samples of the likelihoods $L_i$ and to estimate the corresponding enclosed prior volume $\xi_i$.

Firstly, set the iteration number $i = 0$, the prior volume $\xi_0 = 1$, and draw $N_\text{live}$ {\it live} samples of the unknown image $x$
from the prior distribution $\pi(x)$. Secondly, remove the sample with the smallest likelihood, say $L_{i+1}$, from the live set and
replace it with a new sample. This new sample is again drawn from the prior, but constrained to a higher likelihood than $L_{i+1}$.

It is necessary to then determine the prior volume $\xi_{i+1}$ enclosed by the likelihood level-set (iso-contour) defined by $L_{i+1}$.  This is estimated in a stochastic manner.  The enclosed prior volume for each step $i$ can be estimated by a shrinkage ratio (random variable)
$t_{i+1}$, i.e. by $\xi_{i+1} = t_{i+1} \xi_i$, where
$t_{i+1}$ follows the distribution\footnote{The probability
	distribution \eqref{eqn:shrink-ratio} is for the largest of $N_\text{live}$ samples drawn uniformly from the interval [0, 1].
	This follows since the parameter $x$ is uniformly sampled from the prior $\pi(x)$ and $\{\xi_{i}\}$ are uniformly distributed
	(by the relation $\text{d} \xi = \pi(x) \text{d} x$). }
\begin{align} \label{eqn:shrink-ratio}
	{p} (t) = N_\text{live} t^{N_\text{live}-1}.
\end{align}

Repeat the above step (removing the sample with the smallest likelihood and estimating the updated prior volume) until the entire prior volume (and the nested shells of likelihood) has been traversed.
We finally obtain $\{L_i\}$ and $\{\xi_i\}$ which can then be used to compute the marginal likelihood by \eqref{eqn:baye-evi-compt}.
Moreover, we also simultaneously obtain a set of samples of the parameter $x$ comprising all the discarded (dead) samples
and the $N_\text{live}$ final live samples, which can be used for posterior parameter inferences (refer to Section \ref{sec:NS-post-inf} for further detail).

The volume prior at step $i$ of the nested sampling algorithm, is $\xi_{i} = \prod_{k=1}^i t_{k}$; recall that $t_k$ is the shrinkage ratio and
is independently distributed following the probability density function given in \eqref{eqn:shrink-ratio}.
Since the mean and standard deviation of $\log t$ are respectively
\begin{equation}
	E(\log t) =  - 1/N_\text{live} \quad \text{and} \quad  \sigma (\log t) = 1/N_\text{live},
\end{equation}
we have
\begin{equation}
	\log \xi_i \approx  - i/N_\text{live} \pm \sqrt{i}/N_\text{live}.
\end{equation}
Ignoring uncertainty, one thus takes
\begin{equation}
	\xi_i =  \exp (- i/N_\text{live}).
\end{equation}

A convergence criteria for the nested sampling algorithm should be adopted. Terminating the algorithm too
early or late should be avoided to ensure the marginal likelihood is estimated accurately without unnecessary additional computational cost.  One stopping criterion is that the difference in marginal likelihood estimates
between two iterations falls below a predefined threshold, while another is to ensure a sufficient number of
dead samples is used.

The pseudo code for the nested sampling algorithm is given in Algorithm \ref{alg:nested-sampling}.
Observe that the most challenging task in the nested sampling algorithm is drawing samples from the prior with the hard constraint that samples lie within $\Omega_{L_i}$, i.e.\ within the space defined by the likelihood level-set (see {lines 8--10} in Algorithm \ref{alg:nested-sampling}). This constrained sampling step is relatively easy in small problems but can become very computationally challenging as problem dimension increases. As a result, nested sampling is usually restricted to problems of moderate size.

%-----------------
\begin{algorithm}[h]
	\caption{Nested sampling algorithm}
	\label{alg:nested-sampling}

	{\bf Initialization:} Data $Y$. Set ${\cal Z} = 0$, $\xi_0 = 1$ and $i = 0$. Draw $N_\text{live}$ samples
	$\{x_n\}_{n=1}^{N_\text{live}}$ from the prior distribution $\pi(x)$ in
	the prior space $\Omega$. \\
	{\bf Output:} Evidence ${\cal Z}$ and posterior probabilities $\{p_i\}$.

	\vspace{0.05in}

	{\bf for} $i = 1,\ldots,$ until the stopping criterion reached

	\hspace*{0.1in} - Find the lowest likelihood, say $L_i$, in the set of live samples.

	\hspace*{0.1in} - Compute weight $w_i = (\xi_{i-1} - \xi_{i+1})/2$, where $\xi_i = \exp (- i/N_\text{live})$.

	\hspace*{0.1in} - Update evidence by ${\cal Z} = {\cal Z} + L_i w_i$.

	\hspace*{0.08in} - Draw a new sample from the prior distribution $\pi(x)$ in the restricted parameter space $\Omega_{L_i}$,
	and replace the individual sample associated with the lowest likelihood $L_i$ in the set of live samples.

		{\bf end for} \\
	Update the evidence by ${\cal Z} = {\cal Z} + \frac{w_{i+1}}{N_\text{live}}\sum_{n=1}^{N_\text{live}} {\cal L} (x_n)$. \\
	Compute the posterior probability for each individual sample $p_i = {L_i w_i}/{\cal Z}$.

\end{algorithm}
%-----------------

%-------------------------------------------
\subsection{Error estimation}\label{sec:NS-evidence-error}
%-------------------------------------------

If the prior volumes $\{\xi_i\}$ considered in the discretised integral \eqref{eqn:baye-evi-compt} used to evaluate the marginal likelihood
could be assigned exactly, then
the only error in the estimate of the marginal likelihood would be due to the discretisation of the integral,
which is trivially ${\cal O} (1/{N}^2)$ and negligible when $N$ is sufficiently large. However, since the shrinkage ratio $t_i$ is generated randomly, each prior
volume $\xi_i$ is then assigned approximately, which tends to overwhelm the error brought by
the discretisation of the integral and will therefore cause the dominant source of
uncertainty in the final computed evidence ${\cal Z}$.
This uncertainty, fortunately, can be estimated easily. We recall below the error estimation scheme presented in \citet{S06} using the entropy of the prior volumes. This approach is highly efficient since it does not require any additional sampling.

Let ${\cal P}(\xi) = {\cal L} (\xi)/{\cal Z}$ be the posterior distribution regarding the prior volume $\xi$.
Then the negative relative entropy $H$ can be defined as
\begin{equation} \label{eqn:entropy-h}
	H =  \int {\cal P}(\xi) \log [{\cal P}(\xi)] \text{d} \xi
	\approx \sum_{i=1}^{N} \frac{L_i w_i}{\cal Z} \log \left(\frac{L_i}{\cal Z}\right),
\end{equation}
which can be computed directly from the obtained likelihoods $\{L_i\}$, weights $\{w_i\}$ and the evidence ${\cal Z}$.
Following \citet{S06}, the standard deviation of the uncertainty of $\log {\cal Z}$ using the
nested sampling algorithm reads $\sqrt{H/N_\text{live}}$, i.e.,
\begin{equation} \label{eqn:evi-error}
	\log {\cal Z} =  \log \left( \sum_{i=1}^{N} {L_i w_i} \right) \pm \sqrt{\frac{H}{N_\text{live}}}.
\end{equation}
{In \cite{Chopin2010}, it is established that, under some regularity conditions, the approximation error is asymptotically Gaussian in the limit $N \rightarrow \infty$ and vanishes at the usual Monte Carlo rate $\mathcal{O}(N^{-1/2})$. Moreover, the error scales approximately linearly with the model dimension $d$.}
%-------------------------------------------------------------------

%-------------------------------------------------------------------
\section{Proximal nested sampling framework}\label{sec:method}
%-------------------------------------------------------------------
The main difficulty in applying nested sampling to large inverse problems is to efficiently simulate from the prior distribution subject to a hard likelihood constraint. More precisely, at iteration $i$, the samples from the prior are constrained to the region $\Omega_{L_i}$ defined by the likelihood level-set corresponding to $L_i$ (i.e.\ where a new sample must have a likelihood value greater than $L_i$ at  iteration $i$).

In this section we present our proposed \textit{proximal nested sampling} method to address this challenging constrained sampling problem. Moreover, the proximal nature of the sampling method ensures that non-differentiable distributions, such as popular sparsity-promoting priors involving the $\ell_1$ norm, are supported.
We first present the methodology of proximal nested sampling for arbitrary log-concave distributions of the form \eqref{eqn:baye-thm}. Explicit forms of proximal nested sampling for common choices of priors and likelihoods in imaging sciences are presented in Section~\ref{sec:pns-explicit}.

%-------------------------------------------
\subsection{General constrained sampling problem} \label{sec:problem-setup}
% -------------------------------------------
Following \eqref{eqn:baye-thm} and adopting the notation of Section \ref{sec:NS}, assume that the prior and the likelihood are of the form \mbox{$\pi(x) = \text{exp}(-f(x))$} and \mbox{${\cal L}(x) = \text{exp}(-g(x))$}, where $f$ and $g$ are convex l.s.c.\ (lower semicontinuous) functions on $\Omega$.

We consider sampling from the prior $\pi({x})$, such that ${\cal L}(x) > L^*$ for some generic likelihood value $L^* > 0$.
Let $\iota_{{L}^*}(x)$ and $\chi_{{L}^*}(x)$ be the indicator function
and characteristic function, respectively, defined as
\begin{equation}
	\iota_{{L}^*}(x) =
	\begin{cases}
		1, & {\cal L} (x) > {L}^*, \\
		0, & \text{otherwise},
	\end{cases}
	\quad \text{and} \quad
	\chi_{{L}^*}(x) =
	\begin{cases}
		0,       & {\cal L} (x) > {L}^*, \\
		+\infty, & \text{otherwise}.
	\end{cases}
\end{equation}
Since $\log$ is monotonic,  ${\cal L} (x) > {L}^*$ is equivalent to
$g(x) < \tau$, where
\begin{equation} \label{eqn:tau-likelihood}
	\tau = -\log{L}^*.
\end{equation}
Let ${\cal B}_{\tau} := \{x \ \vert \ g(x) < \tau\}$. Then it is apparent that $\chi_{{L}^*}(x)$, as a constraint for $x$,
is equivalent to $\chi_{{\cal B}_{\tau}}(x)$, where
\begin{equation} \label{eqn:l2-ball-cons}
	\chi_{{\cal B}_{\tau}}(x) =
	\begin{cases}
		0,       & x \in {\cal B}_{\tau}, \\
		+\infty, & \text{otherwise}.
	\end{cases}
\end{equation}

Let ${\pi}_{{L}^*} (x) = \pi(x) \iota_{{L}^*}(x)$ represent the prior distribution with the hard likelihood constraint
${\cal L} (x) > {L}^*$. Since $\iota_{{L}^*}(x) = \text{exp}(- \chi_{{L}^*}(x))$, then we have
\begin{align}
	\begin{split}
		\pi_{{L}^*} (x) & = \pi(x) \iota_{{L}^*}(x)  \\
		& = \text{exp}(-f(x)) \text{exp}(- \chi_{{L}^*}(x))  \\
		& = \text{exp}(- [f(x) + \chi_{{L}^*}(x)] ) \\
		& = \text{exp}(- [f(x) + \chi_{{\cal B}_{\tau}}(x)] ).
	\end{split}
\end{align}
Note that taking logarithm of $\pi_{{L}^*} (x)$ reads
\begin{equation} \label{eqn:log-prior-cons}
	-\log \pi_{{L}^*} (x) = f(x) + \chi_{{\cal B}_{\tau}}(x).
\end{equation}
In the following section we introduce our proximal nested sampling algorithm for parameter $x$ to sample from the constrained prior distribution  \mbox{$\text{exp}(- [f(x) + \chi_{{\cal B}_{\tau}}(x)] )$}.

%---------------------
\subsection{Drawing a sample from the constrained prior} \label{subsec:alg-individual-sample}
%---------------------

Sampling distributions over $\Omega$ is usually challenging because of the dimensionality involved. Sampling from the constrained prior \eqref{eqn:log-prior-cons} is particularly difficult because of the hard constraint that $x \in {\cal B}_{\tau}$, encoded in the characteristic function $\chi_{{\cal B}_{\tau}}(x)$. Sampling is further complicated if the log-prior $f(x)$ is not Lipschitz differentiable over $\Omega$ (e.g.\ for non-differentiable sparsity-promoting priors), since high-dimensional sampling methods rely heavily on gradient information. To circumvent these issues we adopt a proximal MCMC approach, which is particularly suitable for high-dimensional distributions that are log-concave but not smooth. More precisely, in a manner akin to \citet{DMP16}, we use the unadjusted Langevin algorithm (ULA) MCMC sampling strategy combined with Moreau-Yosida approximations of non-differential terms, followed by Metropolis Hastings correction step to control the approximations made, as described in \citet{M15}.

Using the ULA iterative formula, for each given $\tau$ (recall that $\tau$ corresponds to a likelihood value
${L}^*$ by $\tau = -\log{L}^*$; see \eqref{eqn:tau-likelihood}), we can generate the following Markov chain
\begin{equation} \label{eqn:nest-tau-ula}
	x^{(k+1)} = x^{(k)} - \frac{\delta}{2} \nabla \bigl[ f(x^{(k)}) + \chi_{{\cal B}_{\tau}}(x^{(k)}) \bigr] + \sqrt{\delta} { w}^{(k+1)},
\end{equation}
where $\delta > 0$ is the step size and ${ w}^{k+1} \sim {\cal N}(0, \mathbf{1}_K)$ (a $K$-sequence of standard Gaussian random variables).

The non-differentiable characteristic function $\chi_{{\cal B}_{\tau}}(x)$ can be approximated by its Moreau-Yosida envelope $\chi^{\lambda}_{{\cal B}_{\tau}}(x)$, with approximation controlled by $\lambda > 0$.  It is straightforward to show that
\begin{equation} \label{eqn:char-myapprox-explicit}
	\chi^{\lambda}_{{\cal B}_{\tau}}(x) = \frac{1}{2\lambda} \| x - x^*\|_2^2 ,
\end{equation}
where $x^*  $ is the closest point in ${\cal B}_{\tau}$ to $x$, given by the projection of
$x$ onto ${\cal B}_\tau$, i.e. {$x^* = \text{proj}_{{\cal B}_\tau}(x) =
	\text{prox}_{\chi_{{\cal B}_\tau}}(x)$}.
Critically, the $\lambda$-Moreau-Yosida envelope is $\tfrac{1}{\lambda}$-Lipschitz differentiable.  Its gradient can be calculated directly from \eqref{eqn:char-myapprox-explicit} or by noting \eqref{eqn:prox-diff}, yielding
\begin{align} \label{eqn:subdiff-smooth-f}
	\nabla \chi^{\lambda}_{{\cal B}_{\tau}}(x)
	= (x - x^*)/\lambda
	=   (x - \text{prox}_{\chi_{{\cal B}_{\tau}}}(x) ) / \lambda.
\end{align}

Replacing the characteristic function by its Moreau-Yosida approximation in \eqref{eqn:nest-tau-ula} , and noting the gradient \eqref{eqn:subdiff-smooth-f}, yields
\begin{equation} \label{eqn:nest-tau-myula-smooth-f}
	x^{(k+1)} =
	x^{(k)}
	- \frac{\delta}{2}\nabla f(x^{(k)})
	- \frac{\delta}{2\lambda} \bigl[ x^{(k)} - \text{prox}_{\chi_{{\cal B}_{\tau}}}(x^{(k)}) \bigr]
	+ \sqrt{\delta} { w}^{(k+1)}.
\end{equation}
When $f(x)$ is differentiable its gradient can be computed directly (we consider the case where $f(x)$ is non-differentiable shortly).
For differential log-priors $f(x)$, \eqref{eqn:nest-tau-myula-smooth-f} provides the general strategy for sampling from the prior subject to the hard likelihood constraint (with a subsequent Metropolis-Hasting step as discussed below).

If the sample $x^{(k)}$ is already in ${\cal B}_\tau$, i.e.\ $x \in {\cal B}_\tau$, the term \mbox{$\bigl [x^{(k)} - \text{prox}^{\lambda}_{\chi_{{\cal B}_{\tau}}}(x^{(k)}) \bigr]$} disappears and the Markov chain iteration simply involves taking a noisy step to descent the gradient.  In contrast, if  $x^{(k)}$ is not in ${\cal B}_\tau$, i.e.\ $x \notin {\cal B}_\tau$, then a step is taken in the direction \mbox{$- \bigl [x^{(k)} - \text{prox}^{\lambda}_{\chi_{{\cal B}_{\tau}}}(x^{(k)}) \bigr]$}, which acts to move the next iteration in the Markov chain in the direction of the projection of $x^{(k)}$ onto the convex set ${\cal B}_\tau$. This term therefore acts to push the Markov chain back into the constraint set ${\cal B}_\tau$ if it wanders outside of it, although due to the Moreau-Yosida approximation of $\chi_{{\cal B}_\tau}$ it does not guarantee the constraint is satisfied (the subsequent Metropolis-Hasting step does guarantee the hard likelihood constraint is satisfied as discussed below).

When $f(x)$ is non-differentiable, it may be approximated by its differentiable Moreau-Yosida envelope $f^\lambda(x)$.  By noting \eqref{eqn:prox-diff}, the gradient of the term involving the sum of the two Moreau-Yosida approximations then reads
\begin{align} \label{eqn:subdiff}
	\nabla( f^{\lambda}(x) + \chi^{\lambda}_{{\cal B}_{\tau}}(x))
	=  (x - \text{prox}^{\lambda}_f(x) ) / \lambda +  (x - \text{prox}_{\chi_{{\cal B}_{\tau}}}(x) ) / \lambda.
\end{align}
Here we have used the same regularisation parameter $\lambda > 0$ for both approximations for notational brevity, although clearly different parameters can be considered for $f^\lambda(x)$ and $\chi^{\lambda}_{{\cal B}_{\tau}}(x)$ if desired.

Replacing in \eqref{eqn:nest-tau-ula} both $f(x)$ and $\chi_{{\cal B}_{\tau}}(x)$ by their Moreau-Yosida approximations, and noting the gradient \eqref{eqn:subdiff}, yields
\begin{equation} \label{eqn:nest-tau-myula}
	x^{(k+1)} = (1- \frac{\delta}{\lambda}) x^{(k)} + \frac{\delta}{2\lambda} \text{prox}^{\lambda}_f(x^{(k)})  +
	\frac{\delta}{2\lambda} \text{prox}_{\chi_{{\cal B}_{\tau}}}(x^{(k)})  + \sqrt{\delta} { w}^{(k+1)}.
\end{equation}
For non-differentiable log-concave priors, \eqref{eqn:nest-tau-myula} provides the general strategy for sampling from the prior subject to the hard likelihood constraint.

To summarise, given a proper initial sample, say $x^{(0)}$, we generate a Markov chain by iteratively applying the Markov kernel \eqref{eqn:nest-tau-myula-smooth-f} if $f$ is Lipschitz differentiable or the regularised surrogate \eqref{eqn:nest-tau-myula} if it is not, which allows drawing samples from the prior that are likely to be within the likelihood iso-contour $L^*$. This is the main challenge in nested sampling.

The Markov chains generated by ULA-type kernels exhibit some bias resulting from the discretisation of the Langevin stochastic differential equation and from the use of Moreau-Yosida regularisations.  This bias can be asymptotically removed by introducing a Metropolis-Hasting correction step to ensure convergence to the required target density.  In detail, at each iteration, a new candidate $x^{\prime}$ generated using formula
\eqref{eqn:nest-tau-myula-smooth-f} or \eqref{eqn:nest-tau-myula} is then accepted with probability
\begin{equation} \label{eqn:MH}
	\text{min}\left\{1, \frac{q(x^{(k)} \vert  x^\prime) \pi_{L^*}(x^\prime)}{q(x^\prime  \vert  x^{(k)}) \pi_{L^*}(x^{(k)})}  \right\},
\end{equation}
where $q(\cdot \vert \cdot)$ is a transition kernel, which we define by a Gaussian related to the ULA random component (following \citealt{M15}), i.e.,
\begin{equation} \label{eqn:kernel}
	q(x^\prime \vert x^{(k)})  \sim \exp\Big(- \frac{\big(x^\prime - x^{(k)} - \frac{\delta}{2} \nabla \log \pi_{L^*}(x^{(k)}) \big)^2}{2\delta}  \Big).
\end{equation}
If the candidate sample $x^\prime$ is outside of ${\cal B}_\tau$, i.e.\ $x^\prime \notin {\cal B}_\tau$, then $\pi_{L^*}(x^\prime)=0$ and according to the Metropolis-Hasting update the candidate will not be accepted, ensuring the hard likelihood constraint is satisfied.

We summarise our proximal technique to draw an individual sample from the prior under the hard likelihood constraint in Algorithm~\ref{alg:proximal-ind-sample}.

%-----------------
\begin{algorithm}[h]
	\caption{Proximal individual sample draw algorithm}
	\label{alg:proximal-ind-sample}
	{\tt ProxSampleDraw}($x^{(0)}$, $L^*$)  \vspace{0.05in} \\
	{\bf Initialization:}  $k = 0$, $K_\text{gap}$. \\
	{\bf Input:} $x^{(0)}$, $L^*$ (starting point of Markov chain and likelihood threshold). \\
	{\bf Output:} Individual sample $x_\text{new}$ fulfilling {the constraint ${\cal L} (x_\text{new}) > {L}^*$}.

	\vspace{0.05in}

	Compute $\tau = -\log{L}^*$.  \\
	{\bf for} $k = 1,\ldots$

	\hspace*{0.1in} - Compute $x^{(k)}$ using the iterative formula \eqref{eqn:nest-tau-myula-smooth-f} if $f$ is differentiable; otherwise \eqref{eqn:nest-tau-myula}.  \\
	\hspace*{0.1in} - Metropolis-Hasting step following \eqref{eqn:MH} to remove the estimation bias. \\
	\hspace*{0.1in} {\bf if } ${\cal L} (x^{(k)}) > {L}^*$  and  $k \ge K_\text{gap}$ \\
	\hspace*{0.25in} break.   \\
	\hspace*{0.1in} {\bf end if }

	{\bf end for} \\
	Set $x_\text{new} = x^{(k)}$.

\end{algorithm}
%-----------------

%---------------------
\subsection{Initialisation from the unconstrained prior} \label{subsec:alg-proxNS}
%---------------------
The initialisation of the nested sampling method is to draw $N_\text{live}$ samples $\{x_n\}_{n=1}^{N_\text{live}}$ from the prior distribution $\pi(x)$ in
the prior space $\Omega$. If the log-prior $f(x)$ is differentiable this may be applied trivially with the ULA iterative formula.  Otherwise $f(x)$ may again be approximated by its Moreau-Yosida envelope and samples from the prior can be generated by the iterative formula
\begin{equation} \label{eqn:nest-tau-prior}
	x^{(k+1)} = (1- \frac{\delta}{2\lambda}) x^{(k)} + \frac{\delta}{2\lambda} \text{prox}^{\lambda}_f(x^{(k)})
	+ \sqrt{\delta} w^{(k+1)}.
\end{equation}
To draw $N_\text{live}$ samples from the prior, it is necessary to first discard initial samples generated before converging on the target prior distribution. Initial samples corresponding to a number of burn-in iterations, say $K_\text{burn}$, are discarded.
Due to correlations between samples and the algorithm's memory footprint, the chain is thinned by discarding a number of intermediate iterations between samples (the chain's thinning factor), say $(K_\text{gap} - 1)$.  That is, only the $K_\text{gap}$-th sample generated by the iterative formula is kept.  Only 1-in-$K_\text{gap}$ samples are stored when $k > K_\text{burn}$ and ${\tt mod}(k - K_\text{burn}, K_\text{gap}) = 0$,
where ${\tt mod} (\cdot, \cdot)$ represents modulus after division. A Metropolis-Hasting step can also
be introduced here to remove the estimation bias.
We summarise the technique for drawing $N_\text{live}$ live samples from the prior in Algorithm~\ref{alg:proximal-prior}.

%-----------------
\begin{algorithm}[h]
	\caption{Proximal algorithm of drawing live samples (from prior) }
	\label{alg:proximal-prior}
	{\bf Initialization:} $N_\text{live}$, $K_\text{burn}$, $K_\text{gap}$, and $x^{(0)}$. \\
	{\bf Output:} $N_\text{live}$ live samples $\{x_n\}_{n=1}^{N_\text{live}}$ (draw from the prior with no constraint).

	\vspace{0.05in}

	{\bf for} $k = 1,\ldots, K_\text{burn}$ \\
	\hspace*{0.1in} - Compute $x^{(k)}$ using the iterative formula \eqref{eqn:nest-tau-prior}.  \\
	{\bf end for}  \\
	$n = 1$; \\
	{\bf for} $k = K_\text{burn}+1,\ldots, K_\text{burn} + N_\text{live} K_\text{gap}$ \\
	\hspace*{0.1in} - Compute $x^{(k)}$ using the iterative formula \eqref{eqn:nest-tau-prior}. \\
	\hspace*{0.1in} - Metropolis-Hasting step to remove the estimation bias. \\
	\hspace*{0.1in} {\bf if}  ${\tt mod}(k - K_\text{burn}, K_\text{gap}) = 0$ \\
	\hspace*{0.25in}  $x_n$ = $x^{(k)}$; $n = n +1$. \\
	\hspace*{0.1in} {\bf end if} \\
	{\bf end for}

\end{algorithm}
%-----------------

%---------------------
\subsection{Proximal nested sampling algorithm}
% ---------------------

After embedding Algorithms \ref{alg:proximal-ind-sample} and \ref{alg:proximal-prior} into Algorithm \ref{alg:nested-sampling} (i.e., the standard nested sampling algorithm),
we obtain our proposed proximal nested sampling algorithm, which is summarised in Algorithm \ref{alg:proximal-nested-sampling}.
Recall that Algorithm \ref{alg:proximal-ind-sample} generates a new single sample from the prior subject to the hard likelihood constraint, which is used to replace the sample with the lowest likelihood value
in the live sample set. We suggest using a sample randomly selected from the live sample set
as a starting
point for Algorithm \ref{alg:proximal-ind-sample}.

{So far we have presented the proximal nested sampling framework in its most general form for arbitrary log-concave distributions}, which is based on the iterative formula \eqref{eqn:nest-tau-myula-smooth-f} or \eqref{eqn:nest-tau-myula} to sample from the constrained prior.  These iterative formula involve computing proximal operators related to the log-prior and likelihood constraint, which we have not yet considered in further detail. In principle computing proximal operators involves solving a minimisation problem, although in many scenarios this can be solved analytically or otherwise efficient iterative algorithms can be used.  In the following section we consider explicit forms of proximal nested sampling for common forms of the prior and likelihood, outlining explicitly how the required proximal operators can be computed.

%-----------------
\begin{algorithm}[h]
	\caption{Proximal nested sampling algorithm}
	\label{alg:proximal-nested-sampling}

	{\bf Initialization:} Data $Y$. Set ${\cal Z} = 0$, $\xi_0 = 1$ and $i = 0$. Using Algorithm \ref{alg:proximal-prior}
	to draw $N_\text{live}$ samples $\{x_n\}_{n=1}^{N_\text{live}}$ from the prior distribution $\pi(x)$ in the prior space $\Omega$. \\
	{\bf Output:} Evidence ${\cal Z}$ and posterior probabilities $\{p_i\}$.

	\vspace{0.05in}

	{\bf for} $i = 1,\ldots,$ until the stopping criterion reached

	\hspace*{0.1in} - Find the lowest likelihood, say $L_i$, in the set of live samples.

	\hspace*{0.1in} - Compute weight $w_i = (\xi_{i-1} - \xi_{i+1})/2$, where $\xi_i = \exp (- i/N_\text{live})$.

	\hspace*{0.1in} - Update evidence by ${\cal Z} = {\cal Z} + L_i w_i$.

	\hspace*{0.1in} - Randomly select a sample, say $x^{(0)}$, from the set of live samples.

	\hspace*{0.1in} - Use Algorithm \ref{alg:proximal-ind-sample}
	to draw a new sample $x_\text{new}=$ {\tt ProxSampleDraw}($x^{(0)}$, $L_{i}$) from the prior distribution $\pi(x)$ in the restricted parameter space $\Omega_{L_i}$, and replace the individual sample $x_{i, \text{low}}$ by the newly drawn sample $x_\text{new}$.

		{\bf end for} \\
	Update the evidence by ${\cal Z} = {\cal Z} + \sum_{n=1}^{N_\text{live}} {\cal L} (x_n) w_{i+1}/N_\text{live}$. \\
	Compute the posterior probability for each individual sample $p_i = {L_i w_i}/{\cal Z}$.

\end{algorithm}
%-----------------

{Before concluding this section, we note that the proposed proximal nested sampling method summarised in Algorithm \ref{alg:proximal-nested-sampling} seeks to provide a Bayesian model selection strategy that is computationally efficient, simple, robust, and easy to deploy, as opposed to a strategy that seeks to deliver optimal performance by using adaptive methods or by leveraging model-specific properties. For example, for some models with favourable factorisation properties, better results would be obtained by replacing ULA by a Gibbs sampler (see e.g. \citealt{Lucka2016}). Similarly, for models that are close to isotropic, one could replace ULA with a proximal Markov kernel derived from the underdamped Langevin SDE, which includes a Hamiltonian term (see e.g. \citealt{Melidonis22}\footnote{Note that we focus on proximal MCMC kernels
		since purely gradient-based MCMC methods based on the Langevin or Hamiltonian dynamics are not directly applicable to the non-smooth models considered in this paper. They might fail to be geometrically ergodic, in which case the nested sampling scheme would also behave poorly (see \citealt{B11} for an example of a nested sampling method based on Hamiltonian dynamics).}).
	Such methods scale more efficiently to large models than the overdamped Langevin method used in this paper, but they are less robust to anisotropy, which is a common feature in Bayesian inverse problems. Moreover, one could also consider using an adaptive MALA kernel with a matrix-valued step-size taking into account second-order properties of the posterior distribution \citep{PereyraIEEE16}.
	Lastly, because the proposed proximal nested sampling method has been specifically designed for large models that are log-concave, it is not equipped with mechanisms to handle multi-modality. For problems involving multi-modality, we would recommend modifying the Markov kernel either by using some form of annealing \citep{N01}, or by using an adaptive importance sampling scheme \citep{MELC17}. However, as mentioned previously, performing model selection for models that are both large and multi-modal is very difficult and remains an important perspective for future work.}

%-------------------------------------------------------------------
\section{Explicit forms of proximal nested sampling} \label{sec:pns-explicit}
%-------------------------------------------------------------------
In the general proximal nested sampling framework presented in Section~\ref{sec:method} we considered arbitrary log-concave terms for the prior and likelihood and did not consider further how to compute the proximal operators related to those terms.  We now exemplify our proposed proximal nested sampling framework with explicit forms for common priors and likelihoods used in high-dimensional signal and image processing problems.  In particular, we outline explicitly how to compute the required proximal operators.

For illustration, we focus on sparsity-promoting priors corresponding to \linebreak \mbox{$f(x) = \mu \|\mathbf{\mathsf \Psi}^\dagger x\|_1$}, where $\mathbf{\mathsf \Psi}^\dagger \in \mathbb{C}^{p \times d}$ represents a sparsifying transform, and
Gaussian likelihoods corresponding to $g(x) = \|y -\mathbf{\mathsf \Phi} x\|_2^2/{2\sigma^2}$, where $y \in \mathbb{C}^m$ denotes measured data, $x \in \mathbb{R}^d$ the underlying parameters, and $\mathbf{\mathsf \Phi} \in \mathbb{C}^{m\times d}$ the measurement operator (model), although other common priors are also considered.  For simplicity, although not essential, we assume $\mathbf{\mathsf \Psi}$ is an orthonormal transformation, i.e.,
$\mathbf{\mathsf \Psi}^{\dagger}\mathbf{\mathsf \Psi} =\mathbf{\mathsf \Psi}\mathbf{\mathsf \Psi}^{\dagger} = I$.

From the iterative forms given in \eqref{eqn:nest-tau-myula-smooth-f}, \eqref{eqn:nest-tau-myula} and \eqref{eqn:nest-tau-prior}, on which our proximal nested sampling framework is based, it is necessary to compute two proximal operators: $\text{prox}^{\lambda}_f(x)$ and $\text{prox}_{\chi_{{\cal B}_{\tau}}}(x)$, related to the prior and likelihood, respectively (recall that the definition of $\chi_{{\cal B}_{\tau}}$ is related to likelihood function $g$; see \eqref{eqn:l2-ball-cons}).
In the following we calculate these two proximal operators for explicit expressions of $f(x)$ and $g(x)$ and
show the corresponding explicit forms of the iterative formulas of \eqref{eqn:nest-tau-myula-smooth-f},
\eqref{eqn:nest-tau-myula} and \eqref{eqn:nest-tau-prior}.

%-------------------------------------------
\subsection{Proximal operator for the prior} \label{sec:alg-prior}
%-------------------------------------------
When $f(x)$ represents a flat prior or $f(x) = \mu \|\mathbf{\mathsf \Psi}^\dagger x\|_2^2$ (Gaussian prior) it is differentiable with gradient
\begin{equation}
	\nabla f(x) = 0 \quad \text{or} \quad \nabla f(x) = 2 \mu\mathbf{\mathsf \Psi}\mathbf{\mathsf \Psi}^\dagger x = 2 \mu x,
\end{equation}
respectively (here we use $\mathbf{\mathsf \Psi}\mathbf{\mathsf \Psi}^\dagger = I$). Obviously, there is no need to use the Moreau-Yosida envelope $\nabla f^{\lambda}(x)$ to approximate $\nabla f(x)$ when $f(x)$ is differentiable.

When $f(x)$ represents a sparsity-promoting Laplacian-type prior $f(x) = \mu \|\mathbf{\mathsf \Psi}^\dagger x\|_1$,
$\forall x^\prime\in \mathbb{R}^d$, we have
\begin{align}
	\begin{split}
		\text{prox}^{\lambda}_f(x^\prime) & = \argmin_{x\in \mathbb{R}^d} \big\{ \mu \|\mathbf{\mathsf \Psi}^\dagger x\|_1 + \|x - x^\prime\|_2^2/2\lambda \big\}   \\
		& = x^\prime +\mathbf{\mathsf \Psi} \left (\text{prox}^{\lambda \mu}_{\|\cdot\|_1} (\mathbf{\mathsf \Psi}^{\dagger} x^\prime) -\mathbf{\mathsf \Psi}^{\dagger} x^\prime \right)  \\
		& = x^\prime +\mathbf{\mathsf \Psi} \left(\text{soft}_{\lambda \mu}(\mathbf{\mathsf \Psi}^{\dagger} x^\prime) -\mathbf{\mathsf \Psi}^{\dagger} x^\prime \right),
	\end{split}
\end{align}
where the second line follows by standard properties of the proximal operator \citep{CP10}
and where $\text{soft}_{\lambda}(x) = (\text{soft}_{\lambda}(x_1), \text{soft}_{\lambda}(x_2), \cdots)$
is the soft-thresholding operator defined by
\begin{equation}
	\text{soft}_{\lambda}(x_i) =
	\begin{cases}
		0,                             & |x_i| < \lambda,  \\
		x_i (|x_i| - \lambda) / |x_i|, & \text{otherwise}.
	\end{cases}
\end{equation}

%-------------------------------------------
\subsection{Proximal operator for the likelihood} \label{sec:alg-likelihood}
%-------------------------------------------
Consider the Gaussian likelihood corresponding to $g(x) = \|y -\mathbf{\mathsf \Phi} x\|_2^2/{2\sigma^2}$. Recall that $\chi_{{\cal B}_{\tau}} (x) = 0$ if $x \in \{x  \ \vert \ g(x) < \tau \}$ and otherwise $\chi_{{\cal B}_{\tau}} (x) = +\infty$. We are to solve
\begin{align}  \label{eqn:l2-ball-min-orig}
	\begin{split}
		\text{prox}^{\lambda}_{\chi_{{\cal B}_{\tau}}}(x^\prime) & =
		\argmin_{x \in \mathbb{R}^d} \big\{  \chi_{{\cal B}_{\tau}} (x) + \|x - x^\prime\|_2^2/2\lambda \big\} \\
		& = \argmin_{x\in \mathbb{R}^d} \big\{  \chi_{{\cal B}_{\tau}} (x) + \|x - x^\prime\|_2^2 \big\} \\
		& = \text{proj}_{\chi_{{\cal B}_{\tau}}}(x^\prime),
	\end{split}
\end{align}
which is a projection onto set ${\cal B}_{\tau}$.

For the case where the measurement operator is the identity, $\mathbf{\mathsf \Phi} = I$, (e.g.\ denoising problems) then problem \eqref{eqn:l2-ball-min-orig} is the projection onto the $\ell_2$ ball with radius $\sqrt{2 \tau \sigma^2}$.  In this case the proximal (projection) operator has closed-form solution
\begin{equation}  \label{eqn:u_i-identity-close-form}
	\text{proj}_{\chi_{{\cal B}_{\tau}}}(x) =
	\begin{cases}
		x,                                                      & \text{if} \  x \in {\cal B}_{\tau}, \\
		\frac{x - y }{\|x - y \|_2} \sqrt{2 \tau \sigma^2} + y, & \text{otherwise}.
	\end{cases}
\end{equation}

For the case where the measurement operator is not the identity, $\Phi \neq I$, problem \eqref{eqn:l2-ball-min-orig} is equivalent to finding an $x\in \mathbb{R}^d$ satisfying
\begin{equation}  \label{eqn:l2-ball-min}
	\min_{x\in \mathbb{R}^d} \big\{  \chi_{ {\cal B}^\prime_{\tau^\prime } } (u) + \|x - x^\prime\|_2^2/2 \big\},
	\quad \text{s.t.} \ \ u =\mathbf{\mathsf \Phi} x,
\end{equation}
where ${\cal B}^\prime_{\tau} := \{z \ \vert \ \|y - z\|_2^2 <  \tau \}$ and $\tau^\prime = 2 \tau \sigma^2$.
Minimisation problem \eqref{eqn:l2-ball-min} can be solved by a variety of different optimisation methods, e.g.\ by the alternating direction method of multipliers (ADMM) and primal-dual
algorithms (see, e.g., \citealt{NS13} and references therein for further  details).
In the following we present detailed procedures for using the ADMM and primal-dual algorithms to solve problem \eqref{eqn:l2-ball-min}.

%---------------------
\subsubsection{Computation using ADMM method} \label{sec:alg-l2-ball-min-admm}
%---------------------
Firstly, the augmented Lagrangian of the minimisation problem \eqref{eqn:l2-ball-min} can be represented as
\begin{equation}  \label{eqn:l2-admm-al}
	\Lambda(x, u, z) :=
	\chi_{ {\cal B}^\prime_{\tau^\prime } } (u)
	+ \frac{1}{2} \|x - x^\prime\|_2^2
	+ \beta z^\dagger(u -\mathbf{\mathsf \Phi} x)
	+ \frac{\beta}{2} \|u -\mathbf{\mathsf \Phi} x \|_2^2,
\end{equation}
for dual variable $z$ and penalty parameter $\beta > 0$. Starting from an initialisation $x^{(0)}, z^{(0)}$, the augmented Lagrangian of \eqref{eqn:l2-admm-al} can be minimised with respect to variables $u$ and $x$ alternatively, while updating the dual value $z$ using the dual ascent method to ensure the constraint $u=\mathbf{\mathsf \Phi} x$ is satisfied for the final solution, i.e.\
\begin{align}
	u^{(i)}   & =  \argmin_{u \in \mathbb{C}^m} \Lambda(x^{(i)},u,z^{(i)}) , \\
	x^{(i+1)} & =  \argmin_{x \in \mathbb{R}^d} \Lambda(x,u^{(i)},z^{(i)}) , \\
	z^{(i+1)} & = z^{(i)} + u^{(i)} -\mathbf{\mathsf \Phi} x^{(i+1)},
\end{align}
which can be rewritten as the following explicit iterative scheme
\begin{align}
	u^{(i)}   & =  \argmin_{u \in \mathbb{C}^M} \Big\{  \chi_{ {\cal B}^\prime_{\tau^\prime } } (u)
	+ \frac{\beta}{2} \|u -\mathbf{\mathsf \Phi} x^{(i)} + z^{(i)} \|_2^2 \Big\} \label{eqn:u_i},                                                                                      \\
	x^{(i+1)} & =  \argmin_{x\in \mathbb{R}^D} \Big\{ \frac{1}{2} \|x - x^\prime\|_2^2 + \frac{\beta}{2} \|u^{(i)} -\mathbf{\mathsf \Phi} x + z^{(i)} \|_2^2 \Big\},  \label{eqn:x_i1} \\
	z^{(i+1)} & = z^{(i)} + u^{(i)} -\mathbf{\mathsf \Phi} x^{(i+1)}. \label{eqn:z_i1}
\end{align}

The solution to problem \eqref{eqn:u_i} has a closed-form expression since it is the projection onto a scaled and shifted $\ell_2$ ball, i.e.,
\begin{equation}  \label{eqn:u_i-close}
	u^{(i)} =
	\begin{cases}
		\mathbf{\mathsf \Phi} x^{(i)} - z^{(i)},                                                          & \text{if} \ \mathbf{\mathsf \Phi} x^{(i)} - z^{(i)} \in {\cal B}^\prime_{\tau^\prime }, \\
		\frac{\Phi x^{(i)} - z^{(i)} - Y }{\|\Phi x^{(i)} - z^{(i)} - Y \|_2} \sqrt{2 \tau \sigma^2} + Y, & \text{otherwise}.
	\end{cases}
\end{equation}
Problem \eqref{eqn:x_i1} is differentiable and so can be solved by gradient descent.  It is straightforward to show that this problem is equivalent to solving the linear system w.r.t.\  $x$
\begin{equation}  \label{eqn:x_i1-linear}
	(\beta\mathbf{\mathsf \Phi}^{\dagger}\mathbf{\mathsf \Phi} + I) x =  x^\prime + \beta\mathbf{\mathsf \Phi}^{\dagger} (u^{(i)} + z^{(i)}),
\end{equation}
which can be solved by using iterative methods, with $(\beta\mathbf{\mathsf \Phi}^{\dagger}\mathbf{\mathsf \Phi} + I)$ positive definite.

The pseudo code to compute the proximal operator, $\text{prox}_{\chi_{{\cal B}_{\tau}}}(x) $, using ADMM
is summarised in Algorithm \ref{alg:l2-ball-prox-solver}.
Various stopping criteria can be considered, such as a maximum iteration number or the relative error of solutions at two consecutive iterations, i.e., $\|x^{(i+1)} - x^{(i)}\|_2 / \|x^{(i)}\|_2$.

%-----------------
\begin{algorithm}[h]
	\caption{ADMM for proximal operator associated with the likelihood  }
	\label{alg:l2-ball-prox-solver}

	{\bf Initialization:} $x^{(0)}, z^{(0)}$. \\
	{\bf Input:} $x$, $L^*$   \\
	{\bf Output:} $x^*$ (the value of $\text{prox}
		_{\chi_{{\cal B}_{\tau}}}(x) $).

	\vspace{0.05in}

	Compute $\tau = -\log{L}^*$, and form $\chi_{{\cal B}_{\tau}}$.  \\
	{\bf for} $i = 0,\ldots$, until the stopping criterion reached \\
	\hspace*{0.1in} - Compute $u^{(i)}$ by \eqref{eqn:u_i-close};   \\
	\hspace*{0.1in} - Compute $x^{(i+1)}$ by solving \eqref{eqn:x_i1-linear};  \\
	\hspace*{0.1in} - Update $z^{(i+1)}$ by \eqref{eqn:z_i1}.  \\
	{\bf end for} \\
	Set $x^* = x^{(i+1)}$.\\

\end{algorithm}
%-----------------

%---------------------
\subsubsection{Computation using primal-dual method} \label{sec:alg-l2-ball-min-pd}
%---------------------
Alternatively, problem \eqref{eqn:l2-ball-min-orig} can be solved using a primal-dual method.  Note that the problem can be rewritten as
\begin{equation}
	\min_{x\in \mathbb{R}^d} \big\{  \chi_{ {\cal B}^\prime_{\tau^\prime } } (\mathbf{\mathsf \Phi} x) + \|x - x^\prime\|_2^2/2 \big\},
\end{equation}
which is equivalent to the saddle-point problem
\begin{equation} \label{eqn:primal-dual}
	\min_{x\in \mathbb{R}^d} \max_{z \in \mathbb{C}^K} \big\{ z^\dagger\mathbf{\mathsf \Phi} x -
	\chi^*_{ {\cal B}^\prime_{\tau^\prime } } (z) + \|x - x^\prime\|_2^2/2 \big\},
\end{equation}
where $\chi^*_{{\cal B}^\prime_{\tau^\prime }}$ is the convex conjugate of $ \chi_{{\cal B}^\prime_{\tau^\prime }}$.
The saddle-point problem \eqref{eqn:primal-dual} can be solved by alternatively optimising with respect to the primal variable $x$ and the dual variable $z$.
Considering a proximal forward-background step for each alternate optimisation, first for the dual variable $z$ followed by the primal variable $x$, leads to the following iterative scheme
\begin{align}
	z^{(i+1)}       & = \text{prox}_{\chi^*_{{\cal B}^\prime_{\tau^\prime }}} (z^{(i)} + \delta_1\mathbf{\mathsf \Phi} \bar{x}^{(i)}), \label{eqn:pd-z} \\
	x^{(i+1)}       & =  \text{prox}_h (x^{(i)} - \delta_2\mathbf{\mathsf \Phi}^\dagger z^{(i+1)}), \label{eqn:pd-x}                                    \\
	\bar{x}^{(i+1)} & = x^{(i+1)} + \delta_3 (x^{(i+1)} - x^{(i)}), \label{eqn:pd-x_bar}
\end{align}
where $h(x) = \|x - x^\prime\|_2^2/2$, and $\delta_k$, for $k=1, 2, 3$, are algorithm step size parameters.  We next consider how to solve problem \eqref{eqn:pd-z} and \eqref{eqn:pd-x} explicitly.

Problem \eqref{eqn:pd-z} can be solved by
\begin{align}
	z^{(i+1)} & = \text{prox}_{\chi^*_{{\cal B}^\prime_{\tau^\prime }}} (z^{(i)} + \delta_1\mathbf{\mathsf \Phi} \bar{x}^{(i)}) \nonumber \\ \label{eqn:pd-z-sol}  &= z^{(i)} + \delta_1\mathbf{\mathsf \Phi} \bar{x}^{(i)} - \text{prox}_{\chi_{{\cal B}^\prime_{\tau^\prime }}} (z^{(i)} + \delta_1\mathbf{\mathsf \Phi} \bar{x}^{(i)}),
\end{align}
where we have noted the relationship between the proximal operator of the convex conjugate of a function given by \eqref{eqn:con-conj-prox}.
Since ${\cal B}^\prime_{\tau^\prime}$ is an $\ell_2$ ball,
the proximal operator in \eqref{eqn:pd-z-sol} has the closed-form expression
\begin{equation}
	\text{prox}_{\chi_{{\cal B}^\prime_{\tau^\prime }}} (z) =
	\text{proj}_{{{\cal B}^\prime_{\tau^\prime }}} (z) =
	\begin{cases}
		z,                                                      & \text{if} \  z \in {\cal B}^\prime_{\tau^\prime }, \\
		\frac{z - y }{\|z - y \|_2} \sqrt{2 \tau \sigma^2} + y, & \text{otherwise}.
	\end{cases}
\end{equation}

Problem \eqref{eqn:pd-x} is to solve
\begin{equation}
	x^{(i+1)} = \argmin_{x\in \mathbb{R}^d} \big\{ \|x - x^\prime\|_2^2 + \|x - (x^{(i)} - \delta_2\mathbf{\mathsf \Phi}^\dagger z^{(i+1)})\|_2^2 \big\},
\end{equation}
which involves a differentiable objective function and so can be solved analytically, yielding the closed-form solution
\begin{equation}
	x^{(i+1)} =  (x^\prime + x^{(i)} - \delta_2\mathbf{\mathsf \Phi}^\dagger z^{(i+1)})/2. \label{eqn:pd-x-sol}
\end{equation}

The pseudo code to compute the proximal operator, $\text{prox}^{\lambda}_{\chi_{{\cal B}_{\tau}}}(x) $,
using the primal-dual method is summarised in Algorithm~\ref{alg:l2-ball-prox-solver-pd}. The same stopping criterion
as for ADMM in Algorithm~\ref{alg:l2-ball-prox-solver} can also be used for
Algorithm~\ref{alg:l2-ball-prox-solver-pd}.

Note that the main difference between the primal-dual method
and ADMM is that the primal-dual method does not need to solve the linear system in \eqref{eqn:x_i1-linear}. Therefore, the primal-dual method is typically more efficient computationally and is the approach used in the numerical experiments that follow. However, there are specific problems for which the linear system in \eqref{eqn:x_i1-linear} admits a computationally efficient solution and where the ADMM method might be more appropriate.

%-----------------
\begin{algorithm}[h]
	\caption{Primal-dual method for proximal operator associated with the likelihood}
	\label{alg:l2-ball-prox-solver-pd}

	{\bf Initialization:} ${x}^{(0)}, \bar{x}^{(0)}, z^{(0)}$. \\
	{\bf Input:} $x$, $L^*$   \\
	{\bf Output:} $x^*$ (the value of $\text{prox}_{\chi_{{\cal B}_{\tau}}}(x) $).

	\vspace{0.05in}

	Compute $\tau = -\log{L}^*$, and form $\chi_{{\cal B}_{\tau}}$.  \\
	{\bf for} $i = 0,\ldots$, until the stopping criterion reached \\
	\hspace*{0.1in} - Compute $z^{(i+1)}$ by \eqref{eqn:pd-z-sol};   \\
	\hspace*{0.1in} - Compute $x^{(i+1)}$ by solving \eqref{eqn:pd-x-sol};  \\
	\hspace*{0.1in} - Update $\bar{x}^{(i+1)}$ by \eqref{eqn:pd-x_bar}.  \\
	{\bf end for} \\
	Set $x^* = x^{(i+1)}$.\\

\end{algorithm}
%-----------------

%---------------------
\subsection{Explicit iterative formula for drawing samples} \label{sec:alg-sample-express}
%---------------------
We are now in a position to outline the explicit iterative formulas to draw samples for a variety of common priors
using our proximal nested sampling method.

The explicit representations of the iterative equations \eqref{eqn:nest-tau-myula-smooth-f} (for differentiable $f(x)$)
and \eqref{eqn:nest-tau-myula} (for non-differentiable $f(x)$),
which are used in Algorithm~\ref{alg:proximal-ind-sample}
to draw an individual sample from the prior under the hard likelihood constraint, for uniform, Gaussian and Laplacian priors, i.e.\ $f(x)$ constant, $f(x) = \mu\|\mathbf{\mathsf \Psi}^\dagger x\|_2^2$ and $f(x) = \mu \|\mathbf{\mathsf \Psi}^\dagger x\|_1$, respectively, are\begin{align}
	x^{(k+1)} & = (1- \frac{\delta}{2\lambda}) x^{(k)} +
	\frac{\delta}{2\lambda} {x^{*}}^{(k)}  + \sqrt{\delta} w^{(k+1)}, \label{eqn:nest-explicit-flat}                                                                          \\
	x^{(k+1)} & = (1- \frac{\delta}{2\lambda} - {\delta \mu} ) x^{(k)} +
	\frac{\delta}{2\lambda} {x^{*}}^{(k)}  + \sqrt{\delta} w^{(k+1)}, \label{eqn:nest-explicit-gau}                                                                           \\
	x^{(k+1)} & = (1- \frac{\delta}{2\lambda}) x^{(k)} + \frac{\delta}{2\lambda}\mathbf{\mathsf \Psi} \bigl(\text{soft}_{\lambda\mu}(\mathbf{\mathsf \Psi}^{\dagger} x^{(k)})
	-\mathbf{\mathsf \Psi}^{\dagger} x^{(k)} \bigr)
	+ \frac{\delta}{2\lambda} {x^{*}}^{(k)}  + \sqrt{\delta} w^{(k+1)}, \label{eqn:nest-explicit-lap}
\end{align}
where ${x^{*}}^{(k)} = \text{prox}_{\chi_{{\cal B}_\tau}}(x^{(k)})$ is obtained using Algorithm \ref{alg:l2-ball-prox-solver} or \ref{alg:l2-ball-prox-solver-pd}.

Correspondingly, the explicit representations of equation \eqref{eqn:nest-tau-prior}, which is used in Algorithm \ref{alg:proximal-prior}
to draw $N_\text{live}$ initial live samples from the prior distribution $\pi(x)$ in the prior space $\Omega$, are, respectively,
\begin{align}
	x^{(k+1)} & =  x^{(k)} + \sqrt{\delta} w^{(k+1)}, \label{eqn:nest-ini-explicit-flat}                                                         \\
	x^{(k+1)} & = (1 - {\delta \mu} ) x^{(k)} + \sqrt{\delta} w^{(k+1)}, \label{eqn:nest-ini-explicit-gau}                                       \\
	x^{(k+1)} & = x^{(k)} + \frac{\delta}{2\lambda}\mathbf{\mathsf \Psi} \bigl(\text{soft}_{\lambda\mu}(\mathbf{\mathsf \Psi}^{\dagger} x^{(k)})
	-\mathbf{\mathsf \Psi}^{\dagger} x^{(k)} \bigr) + \sqrt{\delta} w^{(k+1)}. \label{eqn:nest-ini-explicit-lap}
\end{align}

{We conclude this section with a brief discussion of the types of priors that the proposed proximal nested sampling method supports. While any prior that is log-concave could be addressed by using proximal nested sampling, we only recommend using the method for priors with proximal operators that are easy to evaluate or to approximate numerically. This is the case for many models used in applied high-dimensional statistics, where inference is often conducted by using convex optimisation algorithms that also require computing proximal operators. For more details about how to evaluate proximal operators, their properties, and lists of functions with known mappings please see \citet{BC11}, \citet{CP10} and \citet[Ch. 6]{NS13}. A library with MATLAB implementations of frequently used proximity mappings is also available online\footnote{ \url{https://github.com/cvxgrp/proximal}}.}

{Moreover, since the proposed proximal nested sampling approach was specifically designed for models that are log-concave and with Bayesian imaging applications in mind, we anticipate that it will be mostly used with informative priors designed to regularise and stabilise high-dimensional estimation problems. As explained in \cite{LMCLD22}, the marginal likelihood can be very sensitive to the choice of the prior. Therefore, it is important that the parameters of the prior are chosen carefully. In particular, we expect that proximal nested sampling will be used in combination with empirical Bayesian strategies that automatically adjust the parameters of the prior by maximum marginal likelihood estimation (see e.g. \citealt{Vidal2020}).}

{Furthermore, high-dimensional Bayesian models that are log-concave often result from a careful trade-off between modelling accuracy and computational tractability, and thus they are inherently misspecified (e.g., in the case of Bayesian imaging applications, one would not expect the prior to define a realistic generative model). Consequently, when using proximal nested sampling in this context one is inherently operating in an $\mathcal{M}$-open Bayesian modelling paradigm, where none of the models under consideration are formally ``true''. We refer the reader to \cite{LMCLD22} for more details about performing model selection in this context, as well as for details about prior sensitivity, objectivity, and the use of data-driven priors in Bayesian model selection.}

%-------------------------------------------------------------------
\section{Numerical experiments}\label{sec:numerics}
%-------------------------------------------------------------------
In this section we validate our proposed proximal nested sampling method and demonstrate its utility on
a range of illustrative problems.

We first validate our method on a problem with a Gaussian likelihood and Gaussian prior
where the value of the marginal likelihood (Bayesian evidence) can be computed analytically.
The dimensions of the problem considered range from low to very high, i.e.\ 2 to $10^6$ dimensions.

Following on from this, we demonstrate the effectiveness of the proximal nested sampling method by applying it to two canonical imaging inverse problems, namely image denoising and image reconstruction.
In particular, we demonstrate the use of proximal nested sampling for the principled Bayesian model selection of the sparsifying dictionary, the regularisation parameter (i.e.\ the $\mu$ parameter of the prior) and the appropriate measurement operator when it may be misspecified.
Furthermore, as mentioned already, as a by-product the samples obtained by nested sampling approaches can also be used to perform posterior inferences.  This is critical in imaging problems in order to recover point estimates, e.g.\ restored images.  Moreover, alternative forms of uncertainty quantification can also be considered from other posterior inferences, e.g.\ variance estimates and posterior credible regions (see, e.g., \citealt{CPM17}).

%---------------------
\subsection{Implementation and computational resources} \label{sec:simulation}
%---------------------

To perform the numerical experiments presented subsequently, the proximal nested sampling algorithms developed in this article were implemented in MATLAB.\footnote{{A Python version of the {\tt proxnest} code implementing the proximal nested sampling framework proposed in this article has since been developed and is available at} \url{https://github.com/astro-informatics/proxnest}.}
The numerical experiments performed to compute the marginal likelihood for
low-dimensional problems (i.e., dimensions less than $200$) were run on a Macbook laptop with an i7 Intel CPU and
memory of 16 GB.
A high-performance workstation, with 24 CPU cores, x86 64 architecture and 256 GB memory,
was used for high-dimensional problems.

%---------------------
\subsection{Validation in high dimensions} \label{sec:evidende-calc}
%---------------------
We first consider the validation of the proximal nested sampling method. For ease of validation,
we consider the prior potential $f(x) =  \mu \|\mathbf{\mathsf \Psi}^\dagger x\|_2^2$,
with $\mu = 1/2$, $\mathbf{\mathsf \Psi} = I$, and the likelihood potential $g(x) = \|y -\mathbf{\mathsf \Phi} x\|_2^2/{2\sigma^2}$, with $\sigma = 1$, $\Phi = I$.
For this setting, we have a closed-form solution of the marginal likelihood value (see Appendix for further details). Test data $y\in \mathbb{R}^d$ are generated by
\begin{equation} \label{eqn:valid-generate-y}
	y = x + w,
\end{equation}
where $x$ is an $d$-dimensional vector of uniformly distributed random numbers in $[0, 1]^d$,
and $w$ is an $d$-dimensional vector of normally distributed random numbers.
Note that the underlying model used to generate the mock data does not match the prior $\pi$ used here,
but that is fine for validation of the calculation of the marginal likelihood. Also, in imaging setting the prior is never perfectly specified. In the following, we consider increasing dimensions from $d=2$ to $d=10^6$. We separate the test into three parts:
i) small models of dimension from $d=2$ to $d=200$, ii) moderately large models of dimension
from $d=2$ to $d=10^5$, and iii) high dimensional models with $d=10^6$.

\begin{figure*}[!htb]
	\begin{center}
		\begin{tabular}{c}
			\includegraphics[trim={{.0\Lwidth} {.0\Lwidth} {.0\Lwidth} {.0\Lwidth}}, clip, width=75mm, height=55mm]{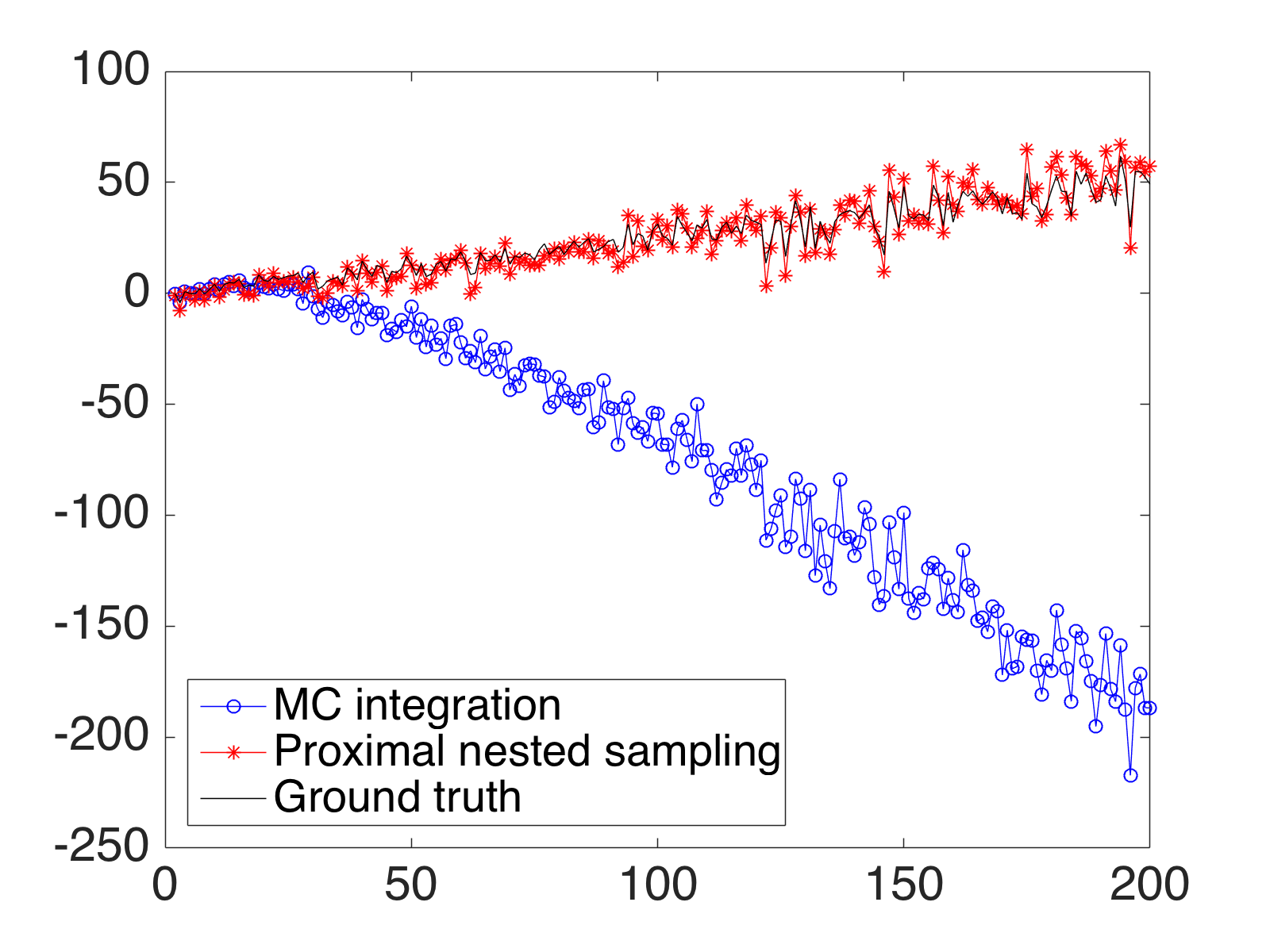}
			\put(-218,48){\rotatebox{90}{ \large $\log(V \times {\cal Z})$  }}
			\put(-125,-5){\large Dimension}
		\end{tabular}
	\end{center}
	\caption{Validation of our proximal nested sampling technique (for dimensions $2$ to $200$) to compute the marginal likelihood (Bayesian evidence) for a scenario where a closed-form solution is accessible.  The logarithm of the unnormalised prior volume ($V$) times the marginal likelihood value (${\cal Z}$) is plotted against the dimensions of the problem considered.
		The blue-circle line, red-asterisk line and the black-solid line show
		the results of MC integration, proximal nested sampling and the ground truth, respectively. We can clearly see that the results computed by proximal nested sampling
		agree with the ground truth well, whereas the result computed by MC integration with $10^{5}$ samples can only achieve acceptable results when the dimension is below $\sim 20$.
		The computation time for the problem with dimension 200 is approximately one minute.
	}\label{Fig:BE-compare}
\end{figure*}

We first test our method for low-dimensional models (i.e., $d < 200$). For our proximal nested sampling method, we use $N_\text{live} = 2\times10^{2}$ live samples and $N = 3\times 10^{3}$ dead samples, with a thinning factor of $10$. We also compare our result with vanilla Monte Carlo (MC) integration where
a uniform prior with integrand $f\cdot g$ is utilised, with the number of samples set to $10^{5}$.
Fig. \ref{Fig:BE-compare} presents the results. Our proximal nested sampling method agrees well with the ground truth,
whereas simple MC integration can only achieve acceptable results when the dimension is small, say $d < 20$. The computation time for the problem with dimension 200 is approximately one minute.

%% ----------------------------------
%% ---------------------------------------

\begin{figure*}[!htb]
	\begin{center}
		\begin{tabular}{c}
			\includegraphics[trim={{.0\Lwidth} {.0\Lwidth} {.0\Lwidth} {.0\Lwidth}}, clip, width=75mm, height=55mm]{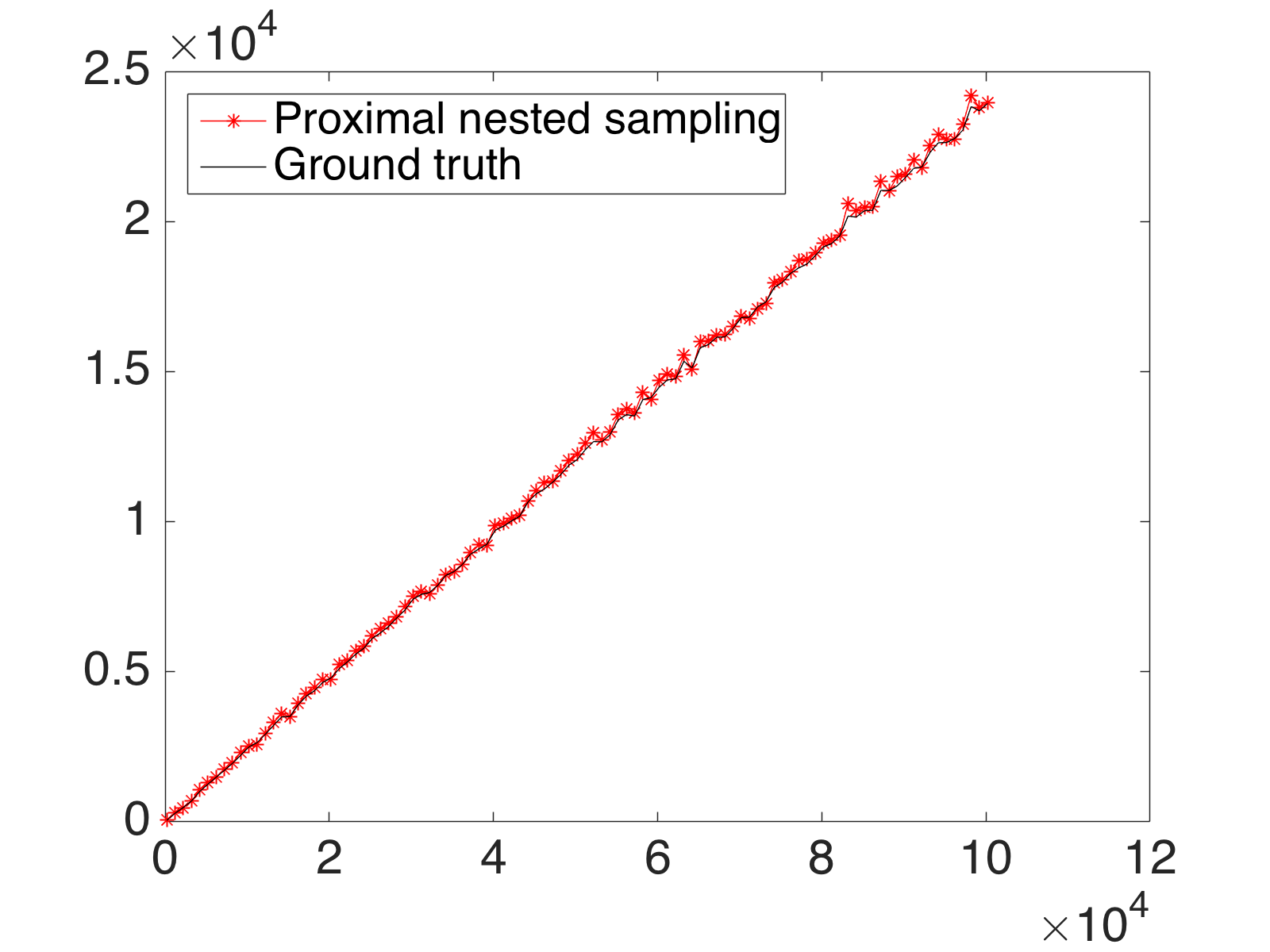}
			\put(-218,48){\rotatebox{90}{ \large $\log(V \times {\cal Z})$ }}
			\put(-125,-5){\large Dimension}
		\end{tabular}
	\end{center}
	\caption{Validation of our proximal nested sampling technique (for dimensions up to $10^5$) to compute the marginal likelihood (Bayesian evidence) for a scenario where a closed-form solution is accessible.  The logarithm of the unnormalised prior volume ($V$) times the marginal likelihood value (${\cal Z}$) is plotted against the dimensions of the problem considered.
		The red-asterisk line and the black-solid line show the results of  proximal nested sampling
		and the ground truth, respectively. We can clearly see that the results computed by proximal nested sampling agrees with the ground truth well.
		The computation time for the problem with dimension $10^5$ is approximately $10$ minutes.
	}\label{Fig:BE-compare-HD}
\end{figure*}

We now test our proximal nested sampling method for high-dimensional cases. Results for dimensions of $y$ up to $10^{5}$ are given in Figure~\ref{Fig:BE-compare-HD},
where we set the number of the live samples $N_\text{live}=10^{3}$ and the number of dead samples $N= 10^4$, with thinning factor $10$ (we do not consider direct MC integration any further since it is already shown to fail for dimensions above $\sim20$).
These results again show that our proximal nested sampling method can achieve results in close agreement with the ground truth. The computation time for the problem with dimension $10^5$ is approximately $10$ minutes.

%% ----------------------------------
%% ---------------------------------------

Finally, we consider dimension $10^6$ as an example to show that our proximal nested sampling method can be pushed to dimensions
much higher than $10^5$. With the same parameters as that used for dimension $10^5$, ten runs were performed for a $10^6$ dimensional setting of the same problem.
The logarithm of the ground truth value was calculated to be $2.3850 \times 10^5$.  The mean of ten runs of proximal nested sampling was computed be to $2.3851 \times 10^5$, with standard deviation $0.0002\times 10^5$.  The result computed by proximal nested sampling is in excellent agreement with the ground truth.
The computation time for each run of the problem with dimension $10^6$ is approximately $30$ minutes.

%---------------------
\subsection{Model selection in image processing} \label{sec:model-select}
%---------------------
We now illustrate the application of proximal nested sampling for Bayesian model selection in imaging problems. In particular, we focus on two canonical problems, image denoising and image reconstruction, with different likelihoods and priors. We emphasise that Bayesian model selection for these imaging problems is not well addressed by existing techniques due to the high dimensions considered (i.e., higher than $10^5$) and the use of general log-concave priors (e.g., like the sparsity promoting Laplace-type priors that include $\ell_1$ terms).
%% ----------------------------------
\begin{figure*}[!htb]
	\begin{center}
		\begin{tabular}{ccc}
			\includegraphics[trim={{.0\Lwidth} {.0\Lwidth} {.0\Lwidth} {.0\Lwidth}}, clip, width=\wwww, height=\hhhh]{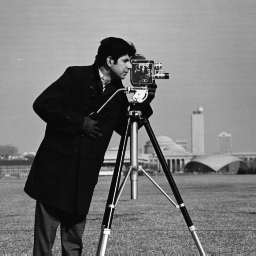}      &
			\includegraphics[trim={{.2\Lwidth} {.17\Lwidth} {.04\Lwidth} {.12\Lwidth}}, clip, width=\ww, height=\hhhh]{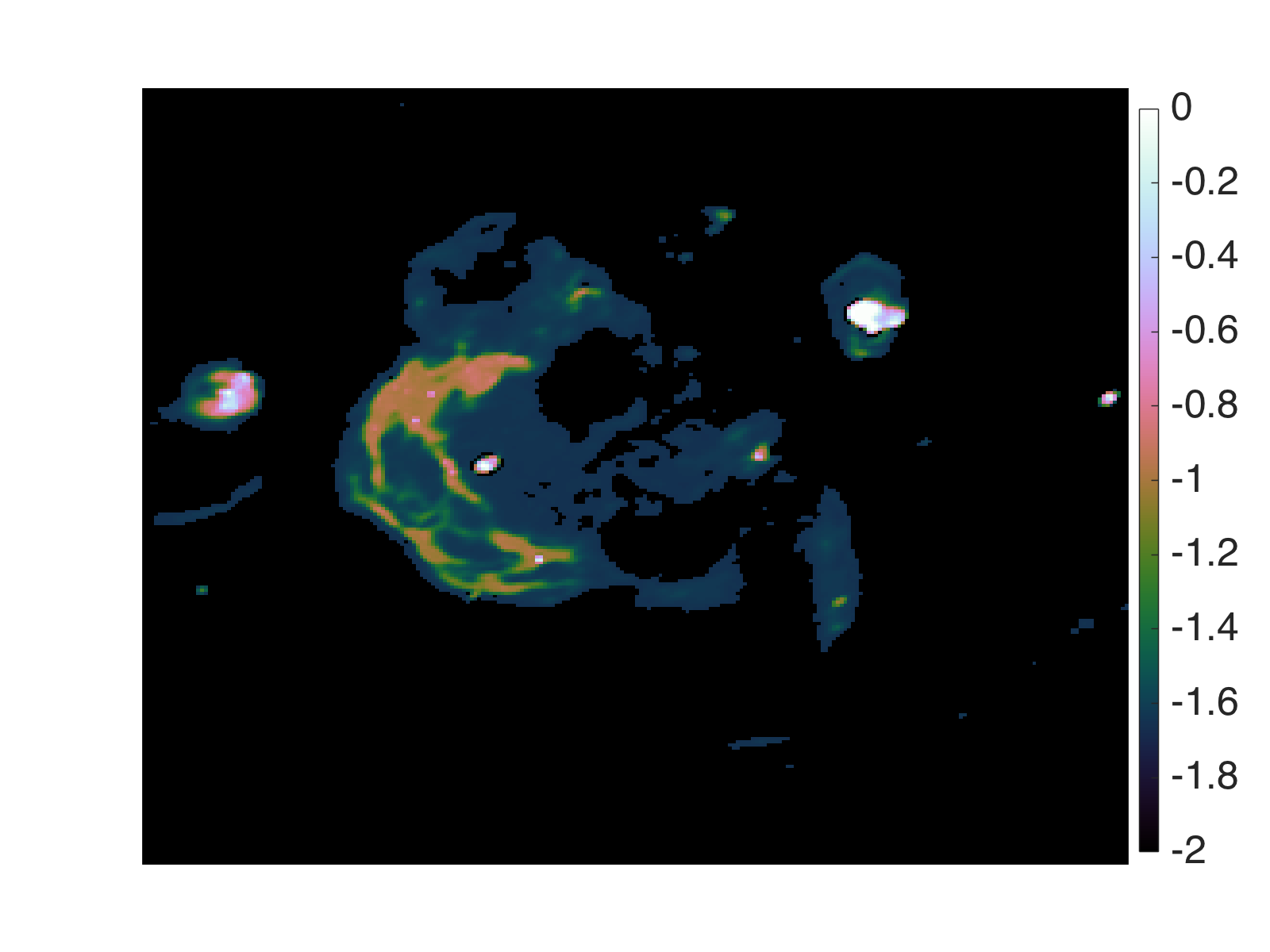} &
			\includegraphics[trim={{.2\Lwidth} {.17\Lwidth} {.04\Lwidth} {.12\Lwidth}}, clip, width=\ww, height=\hhhh]{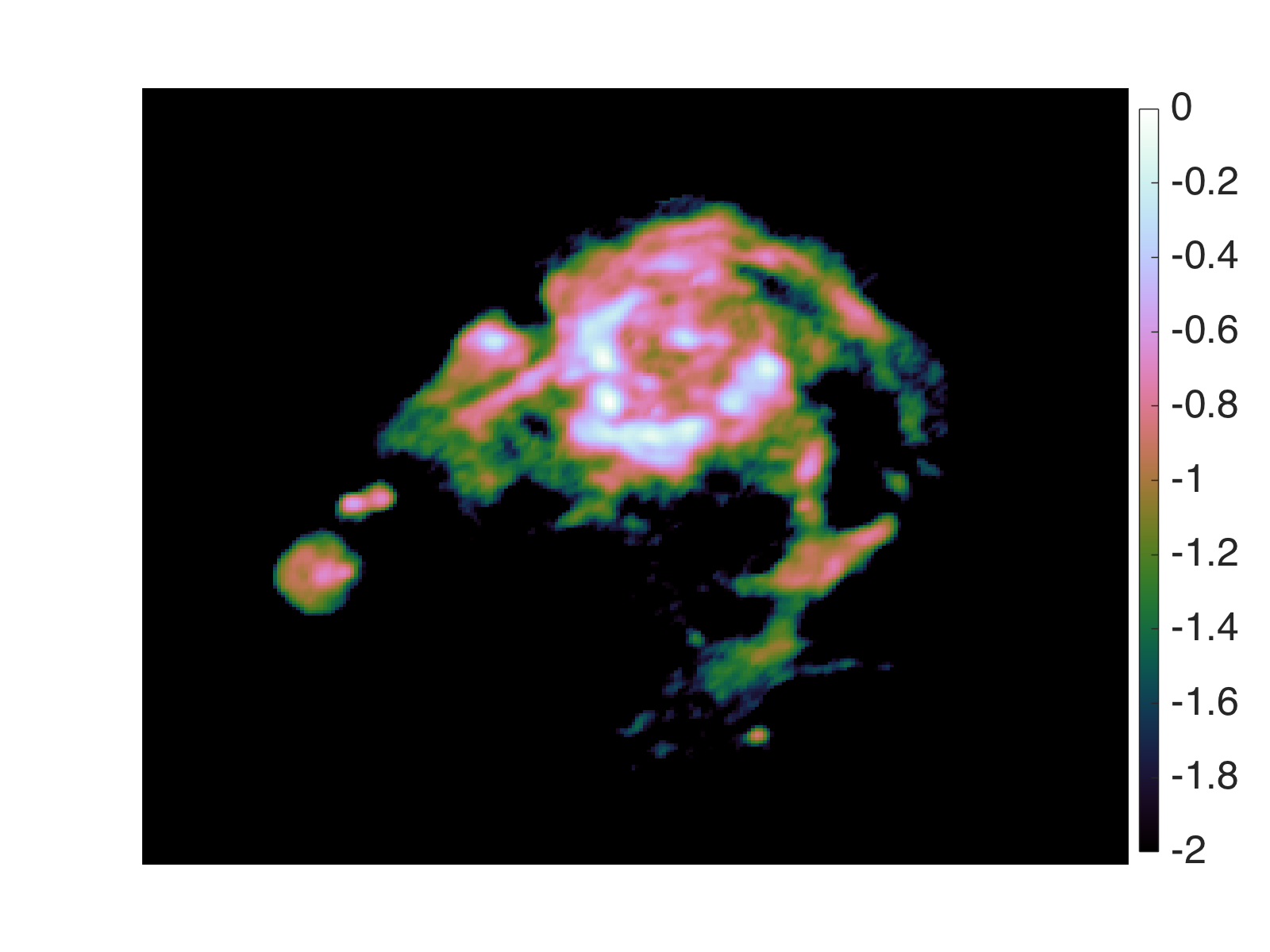}                     \\
			(a) Cameraman                                                                                                                      & (b) W28 & (c) M31
		\end{tabular}
	\end{center}
	\caption{Images used to showcase the use of proximal nested sampling for Bayesian model selection in high-dimensional image processing problems.
		Panel (a): Cameraman grey-scale image; Panels (b)--(c): W28 and M31 radio galaxies normalised to [0, 1] and then shown in {\tt log}10 scale
		(i.e.\ the numeric labels on the colour bar are the logarithms of the image intensity), respectively.
	}\label{Fig:test-image}
\end{figure*}
%% ---------------------------------------

The three images in Fig.~\ref{Fig:test-image} are used in the experiments that follow: Cameraman image, the W28 supernova remnant, and the HI region of the M31 galaxy,  all with size of $256 \times 256$ pixels and with intensities in the range $[0,255]$.
Sparsity-promoting priors (which are not smooth) and Gaussian likelihoods are consider in the following experiments, formed as $f(x) =  \mu \|\mathbf{\mathsf \Psi}^\dagger x\|_1$ and $g(x) = \|y -\mathbf{\mathsf \Phi} x\|_2^2/{2\sigma^2}$, respectively,
where $\mu$, $\mathbf{\mathsf \Psi}$ and $\mathbf{\mathsf \Phi}$ are set to different forms for model selection purposes.

%---------------------
\subsubsection{Prior model selection in image denoising: dictionary selection} \label{sec:model-select-denoise}
%---------------------
For a standard denoising problem we apply proximal nested sampling to select the dictionary $\mathbf{\mathsf \Psi}$ used for the sparsifying transform.
The noisy image $y$ is generated by $y = x + w$,  where $x$ is the ground truth clean image and $w$ is Gaussian noise with zero mean and standard deviation $\sigma = \|x\|_{\infty}10^{-\text{SNR}/20}$, where $\|\cdot\|_{\infty}$ is the infinity norm, and the input signal-to-noise ratio (SNR) is set to 20.
Set $\mathbf{\mathsf \Phi} = I$ in the likelihood \mbox{$g(x) = \|y -\mathbf{\mathsf \Phi} x\|_2^2/{2\sigma^2}$} (i.e., $g(x) = \|y - x\|_2^2/{2\sigma^2}$). We then investigate the influence of different choices for $\mathbf{\mathsf \Psi}$ in the prior term $f(x) =  \mu \|\mathbf{\mathsf \Psi}^\dagger x\|_1$, with $\mu = 10^{5}$.
Specifically, three forms of $\mathbf{\mathsf \Psi}$ are considered, namely the identity ($I$), Daubechies 2 wavelets (DB2), and Daubechies 8 wavelets (DB8).
For the proximal nested sampling method, the number of the live samples $N_\text{live}$ and dead samples $N$ is respectively set to $2\times10^{3}$,
and $4\times10^{4}$ with thinning factor $10^2$, which is sufficient to ensure convergence.

%% ----------------------------------
\begin{figure*}[!htb]
	\begin{center}
		\begin{tabular}{cc}
			\includegraphics[trim={{.0\Lwidth} {.0\Lwidth} {.0\Lwidth} {.0\Lwidth}}, clip, width=\wwww, height=\hhhh]{cameraman.png} &
			\includegraphics[trim={{.23\Lwidth} {.0\Lwidth} {.23\Lwidth} {.0\Lwidth}}, clip, width=\wwww, height=\hhhh]{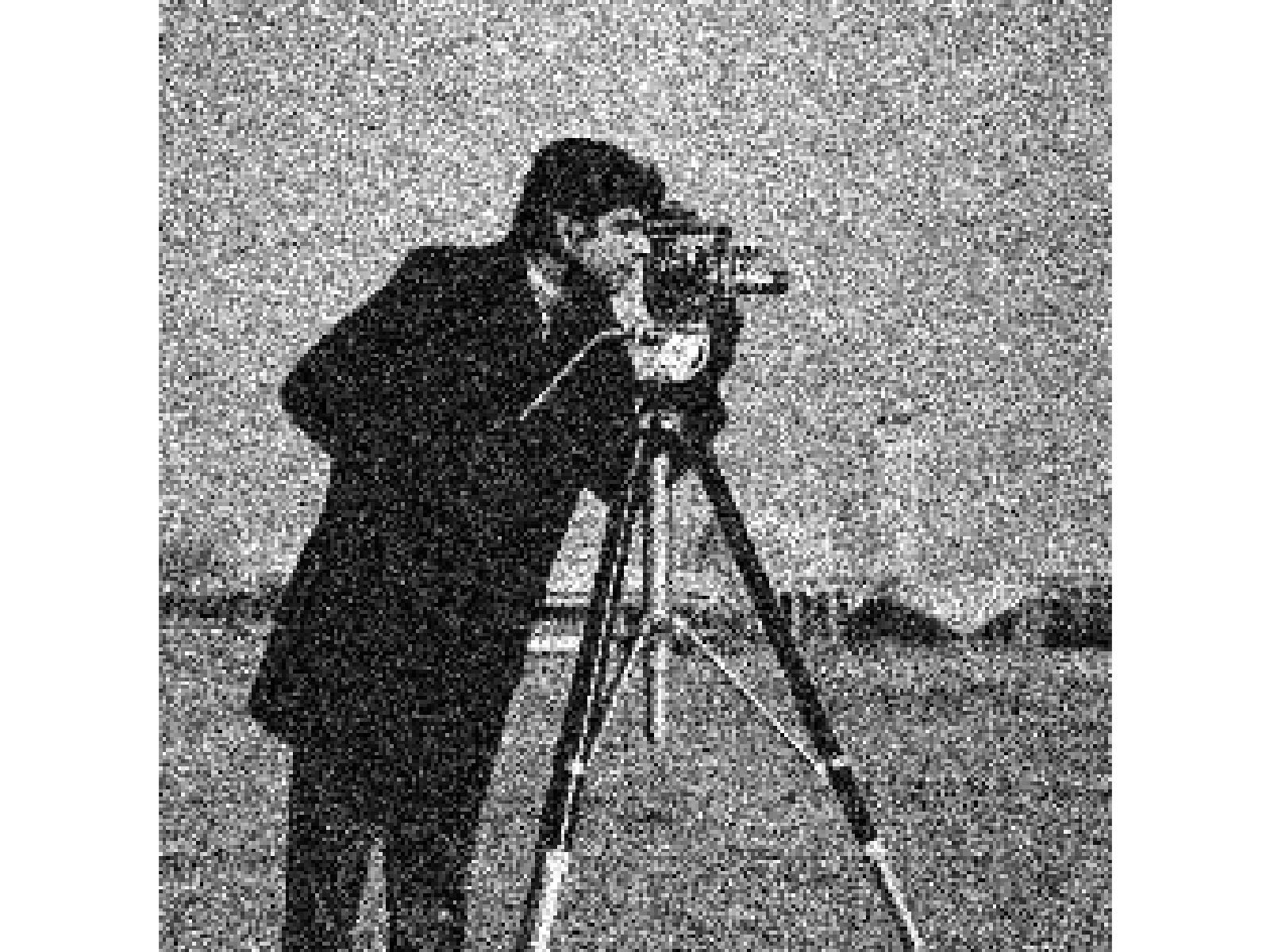}           \\
			(a) Clean image                                                                                                               & (b) Noisy image
		\end{tabular}
		\begin{tabular}{ccc}
			\includegraphics[trim={{.23\Lwidth} {.0\Lwidth} {.23\Lwidth} {.0\Lwidth}}, clip, width=\wwww, height=\hhhh]{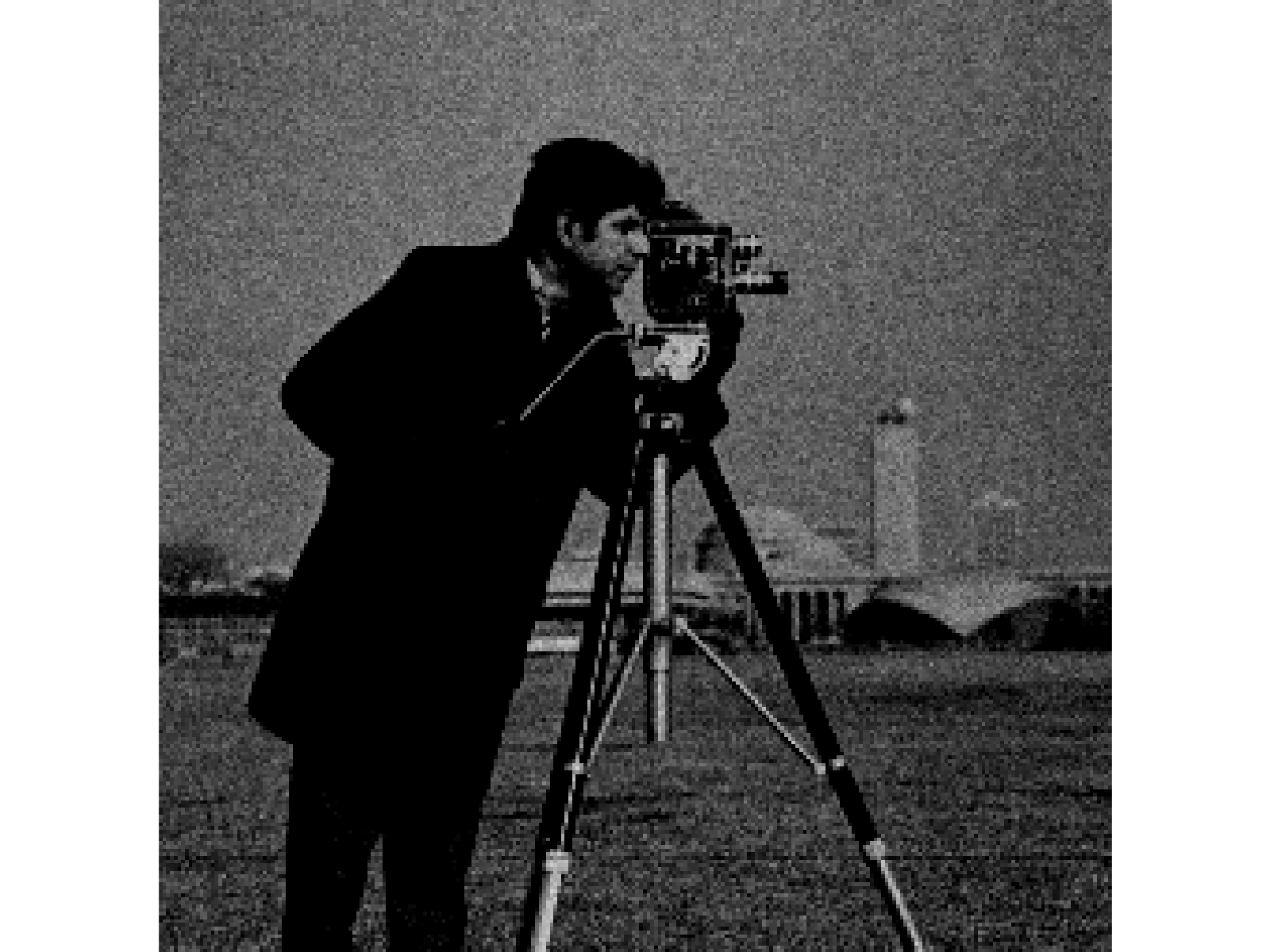} &
			\includegraphics[trim={{.23\Lwidth} {.0\Lwidth} {.23\Lwidth} {.0\Lwidth}}, clip, width=\wwww, height=\hhhh]{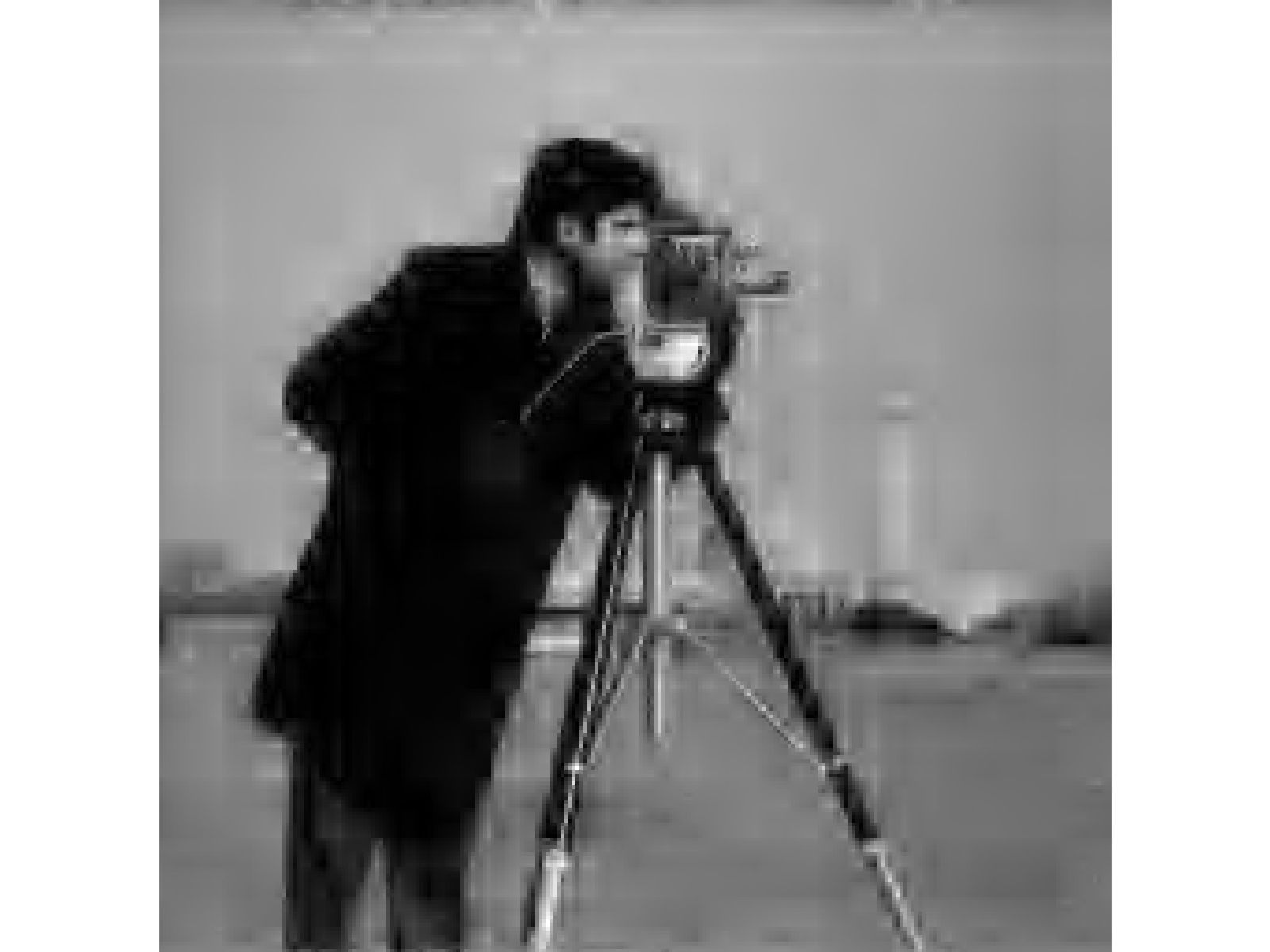}      &
			\includegraphics[trim={{.23\Lwidth} {.0\Lwidth} {.23\Lwidth} {.0\Lwidth}}, clip, width=\wwww, height=\hhhh]{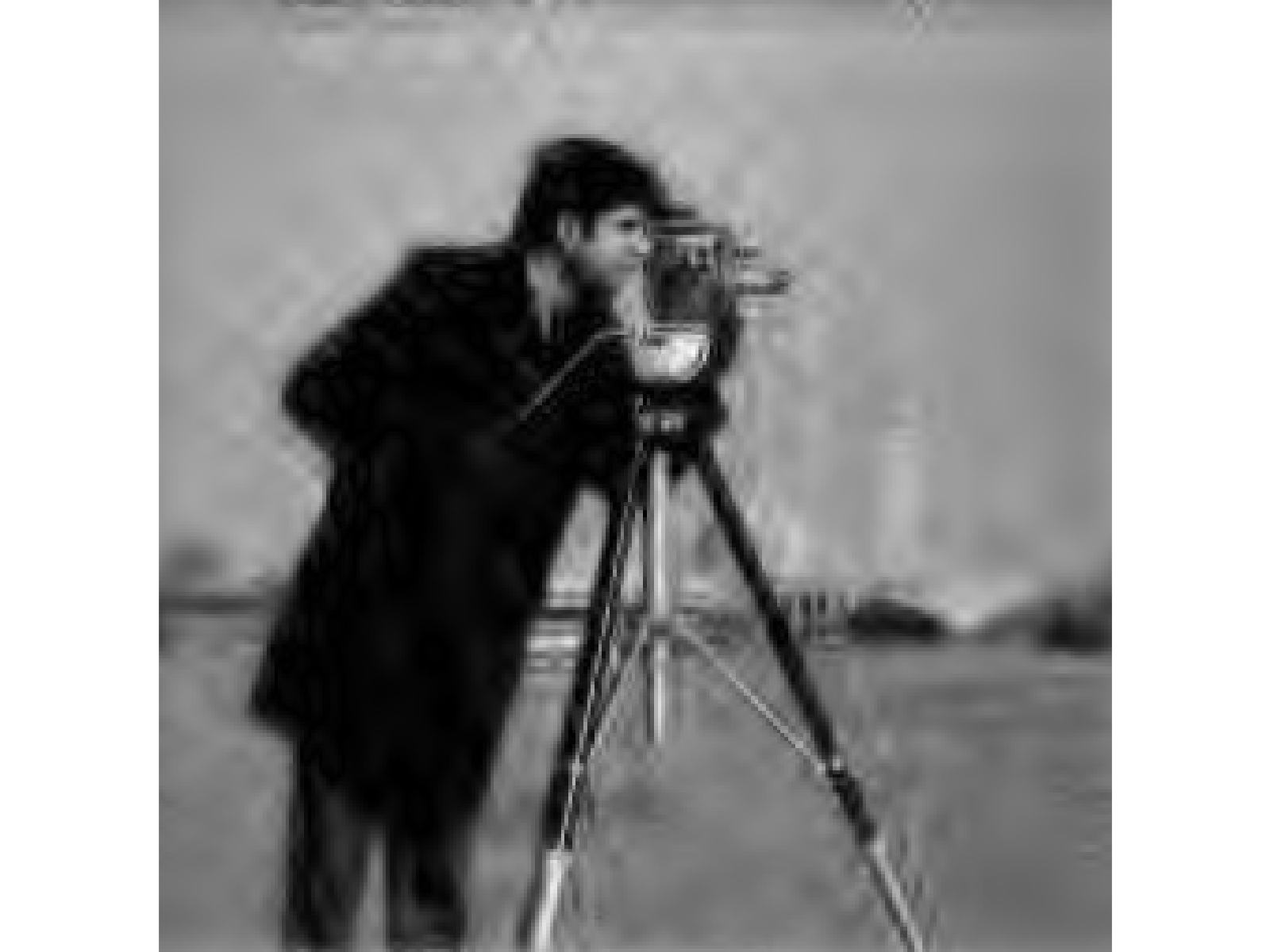}                                                                                             \\
			(c) $\mathbf{\mathsf \Psi} = I$                                                                                                                                                                                                            & (d)  $\mathbf{\mathsf \Psi} = \text{DB2}$ & (e) $\mathbf{\mathsf \Psi} = \text{DB8}$
		\end{tabular}
	\end{center}
	\caption{Dictionary selection for an image denoising problem solved by proximal nested sampling (test image is cameraman).
		First row shows the clean image and noisy image.
		Second row shows the posterior mean images recovered by proximal nested sampling for priors with (sparsifying) transforms $\mathbf{\mathsf \Psi} = I, \text{DB2}$ and $\text{DB8}$, respectively, where the log-prior reads $f(x) = \mu \|\mathbf{\mathsf \Psi}^\dagger x\|_1$. By eye, both DB2 and DB8 wavelets provide superior reconstruction fidelity compared to $\mathbf{\mathsf \Psi}=I$. The model $\mathbf{\mathsf \Psi} = \text{DB2}$ may also be judged to provide slightly superior performance to $\mathbf{\mathsf \Psi}=\text{DB8}$.
	}\label{Fig:test-ms-denoise}
\end{figure*}
%% ---------------------------------------

Fig.~\ref{Fig:test-ms-denoise} presents the posterior means recovered (i.e.\ the reconstructed images) for the three dictionaries considered, i.e.\
for $\mathbf{\mathsf \Psi} = \{ I, \text{DB2}, \text{DB8} \}$.
It is clear that the reconstructed images corresponding to $\mathbf{\mathsf \Psi} = \text{DB2}$ and $\text{DB8}$
are significantly better than that for $\mathbf{\mathsf \Psi} = I$. Moreover, while the difference between the reconstructed images
of the models for $\mathbf{\mathsf \Psi} = \text{DB2}$ and $\text{DB8}$ is small, by eye the image recovered with DB2 may be judged slightly superior.

%% ---------------------------------------
\begin{table}[!h]
	\centering \caption{Marginal likelihood (Bayesian evidence) values computed by proximal nested sampling for Bayesian model selection of the sparsifying dictionary for an image denoising problem (see Fig.~\ref{Fig:test-ms-denoise} for corresponding reconstructed images).  Sparsity-promoting (non-differentiable) priors are considered with (sparsifying) transforms $\mathbf{\mathsf \Psi} = I, \text{DB2}$ and $\text{DB8}$.  Comparing models, Bayesian model selection afforded by proximal nested sampling suggests the model with the DB2 dictionary is superior, followed by DB8, both of which are far superior to the case where $\mathbf{\mathsf \Psi}=I$,  which agrees with the RMSE (root mean square error) values and assessment performed by eye, which require the ground truth to be known.    }
	\begin{tabular}{lcc}
		\toprule
		\multicolumn{1}{c}{Prior}             & $\log {\cal Z}$                & RMSE  \\ \midrule
		$\mathbf{\mathsf \Psi} = I$           & $-6.54 \times 10^4{\pm0.08}  $ & 41.07 \\
		$\mathbf{\mathsf \Psi} = \text{DB2}$  & $-3.06 \times 10^4{\pm0.09}$   & 14.29 \\
		$\mathbf{\mathsf \Psi} = \text{DB8} $ & $-3.09 \times 10^4{\pm0.09}$   & 14.51 \\   \bottomrule
	\end{tabular}
	\label{table-BE-denoise}
\end{table}
%% ---------------------------------------

Table~\ref{table-BE-denoise} presents the calculated marginal likelihood values\footnote{The value of the log marginal likelihoods computed is low (in other words, its absolute value is very high) since the problems we consider are extremely high-dimensional.} for the different sparsifying transforms $\mathbf{\mathsf \Psi}$ selected for the prior.  The root mean square error (RMSE) is also given, where the RMSE gauges the difference between the posterior mean image and the ground truth image. Note that the RMSE cannot normally be computed in
practical problems since the ground truth is not known.  Since for these experiments we know the ground truth the RMSE is a useful measure for comparison purposes.

Table~\ref{table-BE-denoise} shows that the model with $\mathbf{\mathsf \Psi} = I$ possesses the smallest
marginal likelihood value. This implies that for this denoising problem the model with $\mathbf{\mathsf \Psi} = I$ is inferior to models where $\mathbf{\mathsf \Psi}$ is set to DB2 and DB8. Moreover, the marginal likelihood difference between models where $\mathbf{\mathsf \Psi}$ is set to DB2 or DB8 is not dramatic, nevertheless this implies that DB2 is preferred.
These finding inferred by Bayesian model selection agree with the RSME values computed for each model, where the model with $\mathbf{\mathsf \Phi}=\text{DB2}$ is slightly preferred over DB8, and both models with DB2 and DB8 are highly preferred over the model with $\mathbf{\mathsf \Phi}=I$ (recall that in practice it is not possible to compute the RMSE since it requires knowledge of the underlying ground truth).  Furthermore, the model preferences inferred by proximal nested sampling also agree with the assessment of reconstructed image quality by-eye discussed above.
The results obtained are consistent
with common knowledge that it is typically more effective to denoise a natural image using a prior that promotes sparsity in some (sparsifying) transform domain (e.g.\ a wavelet domain)  rather than in the image domain itself.
The computation time for the problem with $\mathbf{\mathsf \Psi} = I$ is approximately $10$ minutes, and for the problem with $\mathbf{\mathsf \Psi} = \text{DB2}$ or $\text{DB8}$ is approximately $60$ minutes.

In high-dimensional settings note that Bayes factors can be very large due to the concentration of probability in high-dimensions, hence it is not meaningful to consider traditional scales for assessing model comparisons such as the Jeffery's scale \citep{NG13}.  Instead, we recommend comparing marginal likelihood values directly.

%---------------------
\subsubsection{Prior model selection in image reconstruction: regularisation parameter selection} \label{sec:model-select-recons-reg}
%---------------------
We now apply proximal nested sampling to a standard reconstruction problem and, firstly, consider the selection of the regularisation parameter $\mu$ defining the width of the prior.  It is typically very challenging to optimally set the regularisation parameter $\mu$, which controls the strength of prior knowledge and plays a key role in reconstruction quality.
Consider noisy observations (noisy measurements)
\begin{equation} \label{eqn:data-y}
	y =\mathbf{\mathsf \Phi} x + w,
\end{equation}
where $w$ again denotes Gaussian noise with zero mean and
$\sigma = \|x\|_{\infty}10^{-\text{SNR}/20}$ (standard deviation),
with SNR set to 30, and $m$ and $d$ are respectively the dimension of $y$ and image $x$.
Consider the prior $f(x) =  \mu \|\mathbf{\mathsf \Psi}^\dagger x\|_1$, with $\mathbf{\mathsf \Psi} = \text{DB8}$, and likelihood $g(x) = \|y -\mathbf{\mathsf \Phi} x\|_2^2/{2\sigma^2}$.
For the reconstruction scenario, $\mathbf{\mathsf \Phi}$ represents the sensing (measurement) operator.
In particular, we consider a measurement model comprising incomplete Fourier measurements (common in radio interferometric and magnetic resonance imaging) defined by the sensing operator \mbox{${\mathbf\mathsf\mathbf{\mathsf \Phi}} = \mathbf{\mathsf M}\mathbf{\mathsf F}$}, constructed from the Fourier transform $\mathbf{\mathsf F}$ followed by a selection mask $\mathbf{\mathsf M}$ which is generated randomly
through the variable density sampling profile \citep{PVW11}.
We consider the scenario where only 30\% of Fourier coefficients are measured, i.e.\ $m = 0.3 d$.  Note that different forms of the mask $\mathbf{\mathsf M}$ result in different sensing operators $\mathbf{\mathsf \Phi}$.

%% ----------------------------------
\begin{figure*}[!htb]
	\begin{center}
		\begin{tabular}{ccc}
			\includegraphics[trim={{.4\Lwidth} {.37\Lwidth} {.04\Lwidth} {.26\Lwidth}}, clip, width=\ww, height=\hhhh]{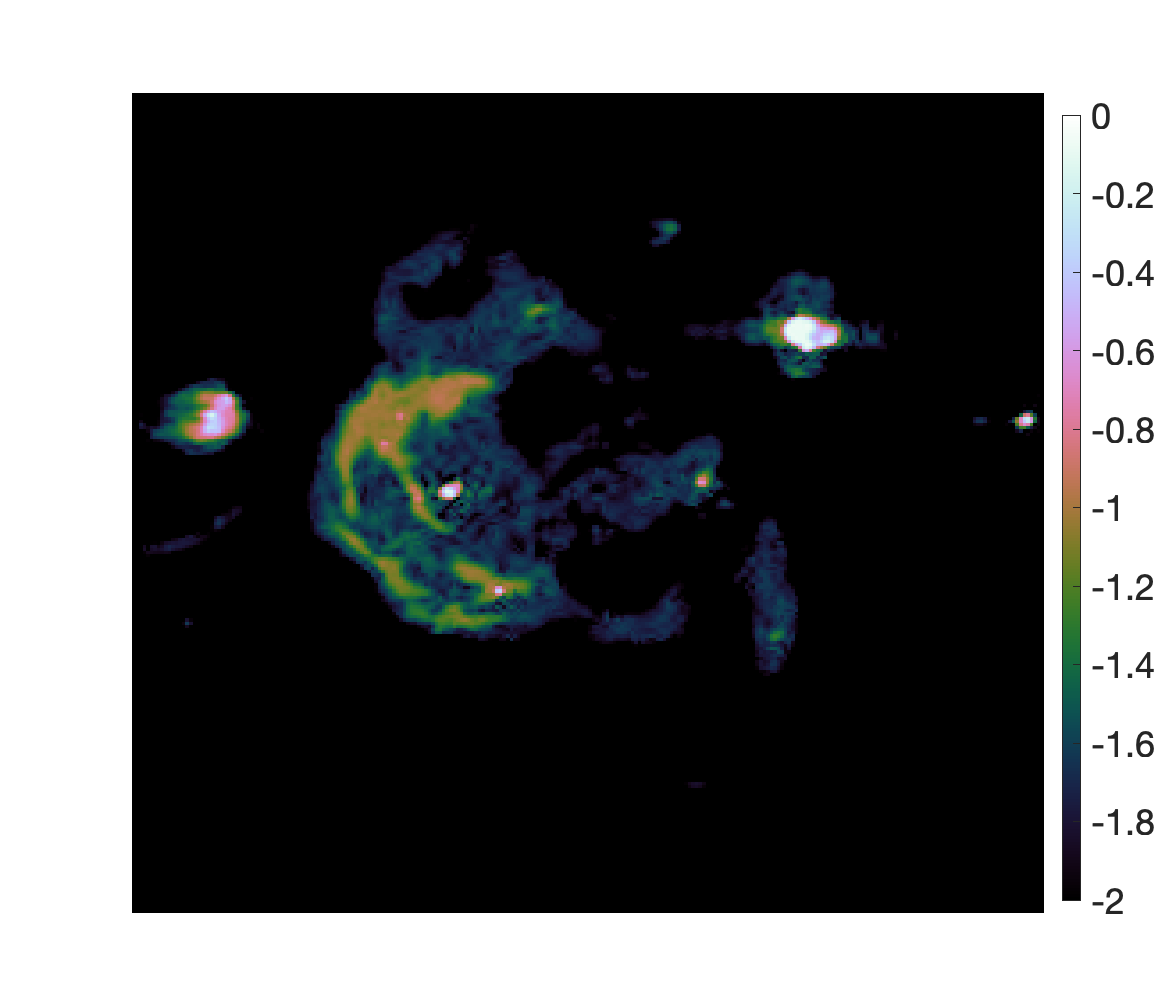}  &
			\includegraphics[trim={{.4\Lwidth} {.37\Lwidth} {.04\Lwidth} {.26\Lwidth}}, clip, width=\ww, height=\hhhh]{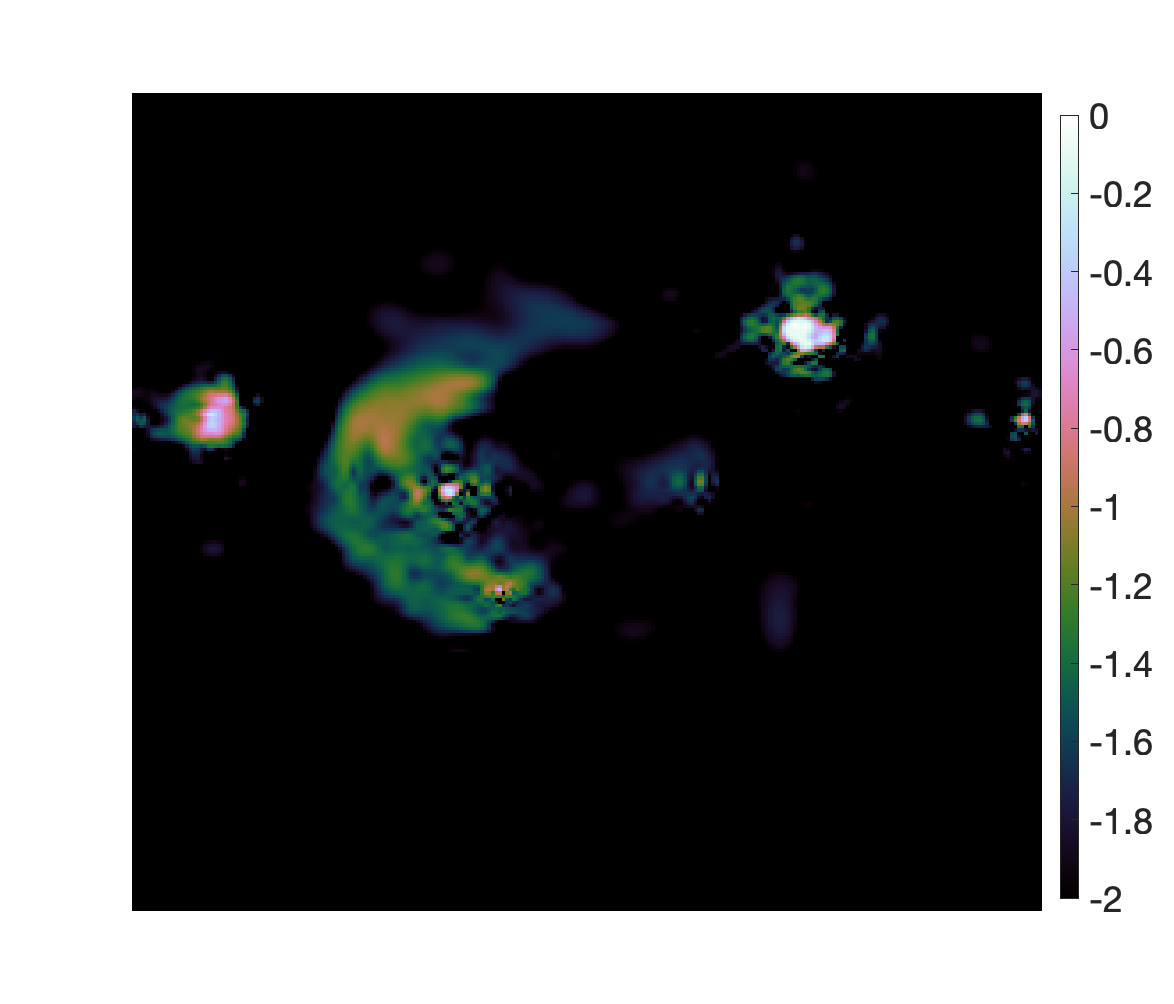} &
			\includegraphics[trim={{.4\Lwidth} {.37\Lwidth} {.04\Lwidth} {.26\Lwidth}}, clip, width=\ww, height=\hhhh]{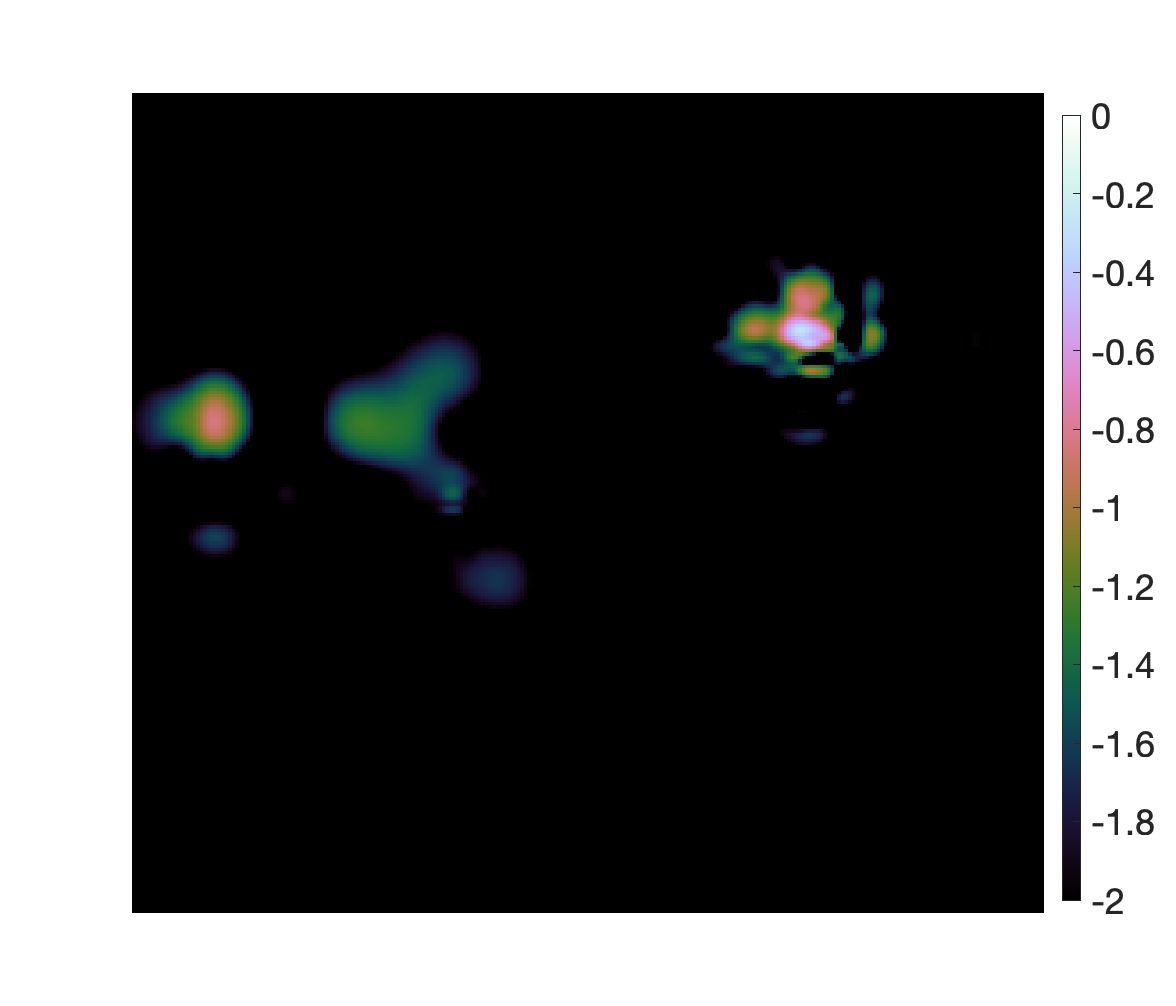}                              \\
			$\mu = 10^6$                                                                                                                                                                                                                          & $\mu = 10^7$ & $\mu = 10^8$
		\end{tabular}
	\end{center}
	\caption{Regularisation parameter selection for an image reconstruction problem solved by proximal nested sampling (test image is W28 radio galaxy).
		Images from left to right are the posterior mean images recovered by proximal nested sampling for
		$\mu$ in the prior definition set to $10^6, 10^7$ and $10^8$, respectively.  The data $y$ are generated by measuring 30\% of noisy Fourier coefficients of the test image.
		On close inspection it may be noticed that reconstruction for model with $\mu = 10^6$ is superior to the one with $\mu = 10^7$, which is superior to the one with $\mu = 10^8$.
	}\label{Fig:test-ms-recons-M31-W28-p}
\end{figure*}
%% ---------------------------------------

%% ---------------------------------------
\begin{table}[!h]
	\centering \caption{Marginal likelihood (Bayesian evidence) values computed by proximal nested sampling for Bayesian model selection of the regularisation parameter $\mu$ for an image reconstruction problem (see Fig.~\ref{Fig:test-ms-recons-M31-W28-p} for corresponding reconstructed images). Prior definition with $\mu$ set to $10^6, 10^7$ and $10^8$, respectively, are considered.  Comparing models, Bayesian model selection afforded by proximal nested sampling suggests the model with $\mu = 10^6$ is superior to the one with $\mu = 10^7$, which is superior to the one with $\mu = 10^8$,  which agrees with the RMSE (root mean square error) values and assessment performed by eye, which require the ground truth to be known. }
	\begin{tabular}{lcc}
		\toprule
		\multicolumn{1}{c}{$\mu$} & $\log {\cal Z}$              & RMSE \\ \midrule
		$10^6$                    & $-2.61\times 10^4{\pm0.09}$  & 1.82 \\
		$10^7$                    & $-5.39\times 10^4{\pm0.09} $ & 2.81 \\
		$10^8$                    & $-2.90\times 10^5{\pm0.09} $ & 6.70 \\  \bottomrule
	\end{tabular}
	\label{table-BE-recons-p}
\end{table}
%% ---------------------------------------

Fig.~\ref{Fig:test-ms-recons-M31-W28-p} presents the posterior means recovered by proximal nested sampling (i.e.\
the reconstructed images) for models with $\mu$ set to $10^6, 10^7$ and $10^8$.
It is difficult to assess the effectiveness of different regularisation parameters by eye, but on close inspection it may be noticed that the model with $\mu = 10^6$ is superior to the one with $\mu = 10^7$, which is superior to the one with $\mu = 10^8$. The computation time for each problem is approximately $150$ minutes.

Table \ref{table-BE-recons-p} presents the marginal likelihood and RMSE values computed for the models with different regularisation parameters $\mu$.
The computed marginal likelihood for the model with $\mu = 10^6$ is larger that the value for the model with $\mu = 10^7$, which is larger than the model with $\mu = 10^8$, suggesting the model with $\mu = 10^6$ is preferred.
The computed marginal likelihoods are consistent with the model preferences obtained by comparing the RMSE of each model and by visual inspection. Recall that both RMSE and visual comparisons can only be performed here where the ground truth is available and cannot be used for model comparison in practice.
In summary, this example demonstrates that our proximal nested sampling method is capable of selecting superior
regularisation parameters for models stemming from high-dimensional inverse problems.

%---------------------
\subsubsection{Measurement model selection in image reconstruction} \label{sec:model-select-recons-meas}
%---------------------
We now apply proximal nested sampling to the same reconstruction problem considered above (i.e.\ image reconstruction from noisy and incomplete Fourier measurements) but focus on the problem of misspecification of the measurement model $\mathbf{\mathsf \Phi}$.
Noisy observations $Y$ are generated by \eqref{eqn:data-y}, measuring 10\% of Fourier coefficients, i.e.\ with  $m = 0.1 d$.

We use the ground truth model $\mathbf{\mathsf M}_\text{truth}$ to simulate observation data $y$.  However, when solving the resulting inverse problem we consider a number of different measurement models, not only the ground truth model $\mathbf{\mathsf M}_\text{truth}$ but also misspecified models $\mathbf{\mathsf M}_\gamma$, where $\gamma > 0$ encodes the level of misspecification.

The method by which the model is misspecified in motivated by radio interferometric imaging.  In radio interferometry, the coordinates of the Fourier coefficients acquired by the telescope are measured in units of (radio) wavelength.  If the wavelength at which observations are made is misspecified, the coordinates of the Fourier coefficients acquired will be scaled.  We model precisely this type of misspecified model here to represent the case where the instrument wavelength is not calibrated accurately.

An incorrectly specified wavelength then simply acts to modify the mask of the ground truth measurement model $\mathbf{\mathsf M}_\text{truth}$.  The misspecified model corresponding to mask $\mathbf{\mathsf{M}}_{\gamma}$, for misspecification parameter $\gamma$, is generated by extending every measured position in $\mathbf{\mathsf{M}}_\text{truth}$ radially.
Specifically, every measured position is extended radially along the line connecting it to the origin to a length of $\gamma d_j$, $j \in \Omega_\text{mask}$, where $\gamma$ is the misspecification scaling factor,
$d_j$ is the distance from the original measured position $j$ to the origin in $\mathbf{\mathsf{M}}_\text{truth}$, and $\Omega_\text{mask}$ is
the set which contains all the measured positions.
It is worth mentioning that the larger the scaling factor $\gamma$, the
larger the distortion of $\mathbf{\mathsf M}_\gamma$ from the ground truth $\mathbf{\mathsf M}_\text{truth}$.  Note also that $\gamma=0$ corresponds to a correctly specified model, i.e.\ $\mathbf{\mathsf M}_{\gamma=0}=\mathbf{\mathsf M}_\text{truth}$.

For proximal nested sampling, the number of the live samples $N_\text{live}$ and dead samples $N$ is respectively set to $2\times10^{3}$
and $3\times10^{4}$ with thinning factor $10^2$, which is sufficient to ensure convergence.
Regularisation parameter $\mu  = 10^8$ is used for these experiments.

%% ----------------------------------
\begin{figure*}[!htb]
	\begin{center}
		\begin{tabular}{ccc}
			\includegraphics[trim={{.2\Lwidth} {.17\Lwidth} {.04\Lwidth} {.12\Lwidth}}, clip, width=\ww, height=\hhhh]{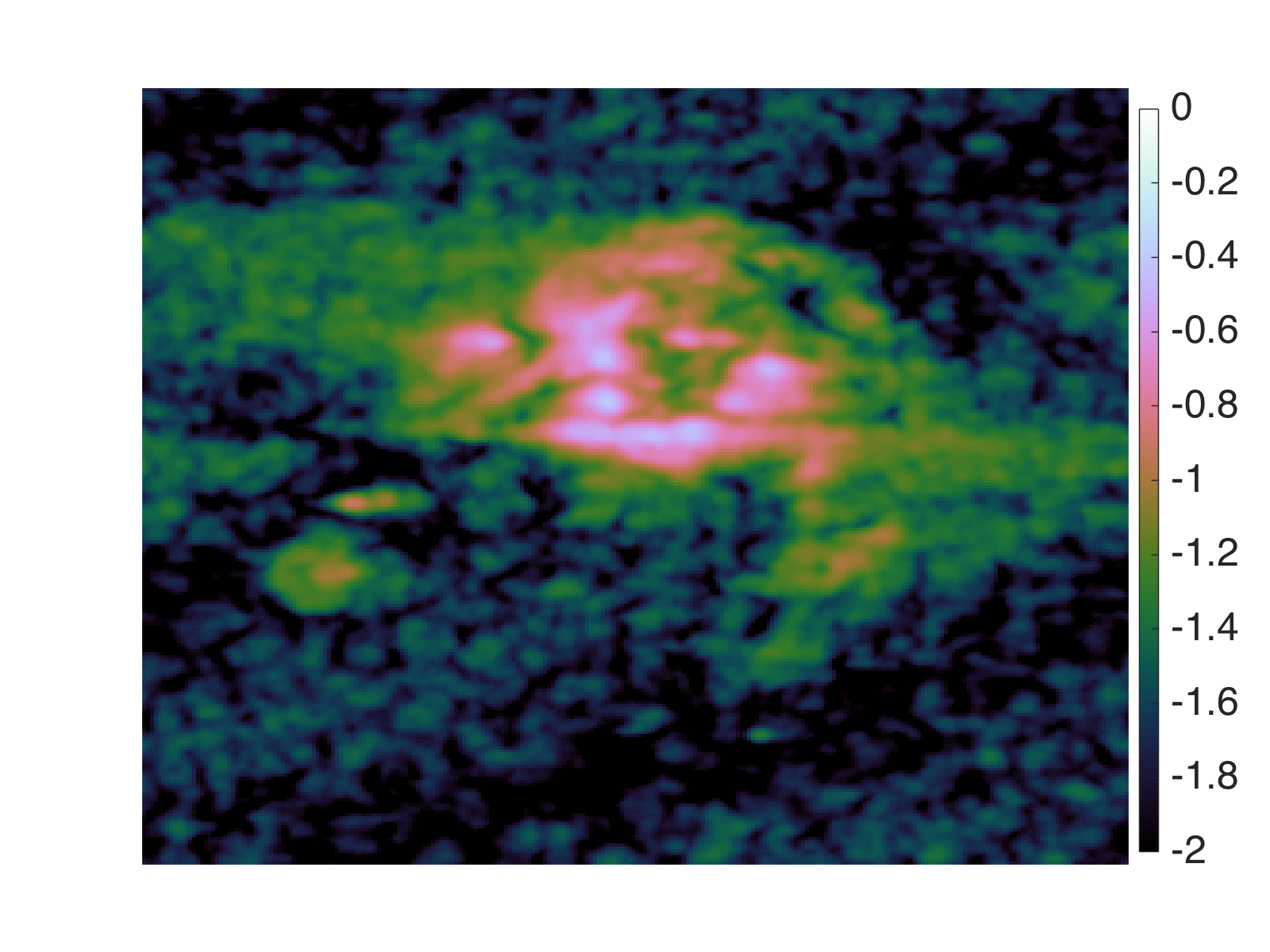}                                                                                                 &
			\includegraphics[trim={{.2\Lwidth} {.17\Lwidth} {.04\Lwidth} {.12\Lwidth}}, clip, width=\ww, height=\hhhh]{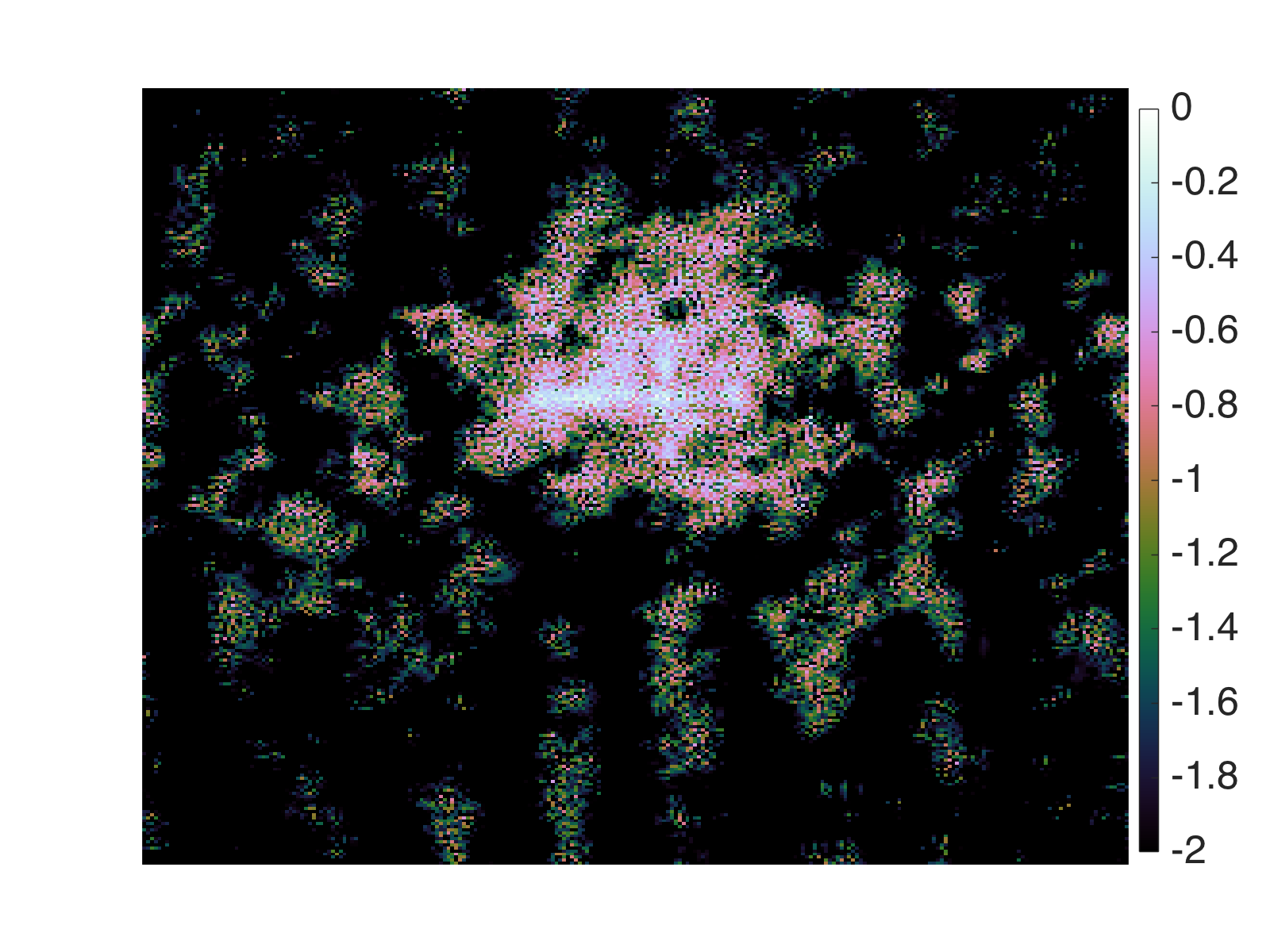} &
			\includegraphics[trim={{.2\Lwidth} {.17\Lwidth} {.04\Lwidth} {.12\Lwidth}}, clip, width=\ww, height=\hhhh]{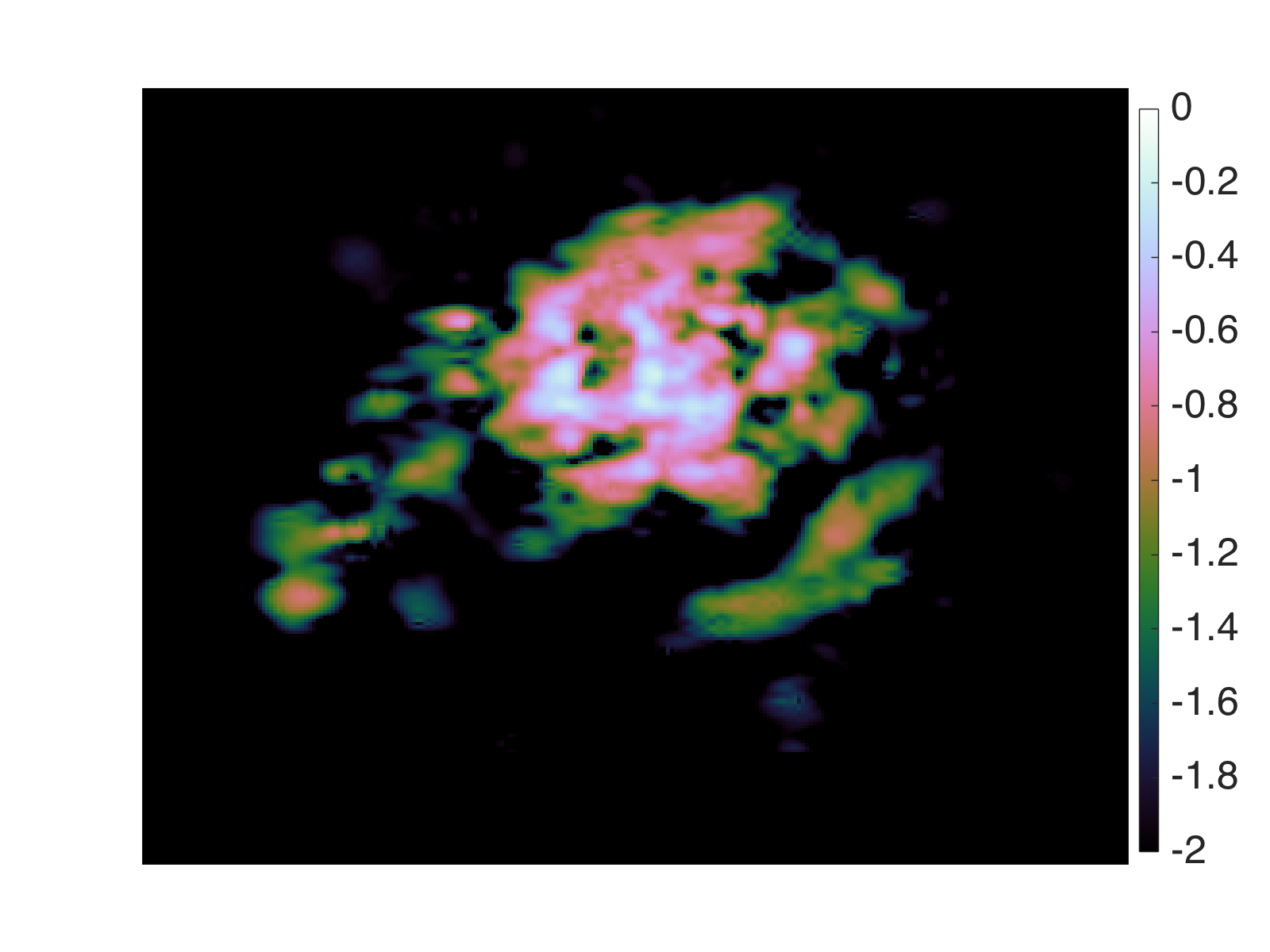}                                                                                                                                                                \\
			(a) Dirty map                                                                                                                                                                                                                      & (b) $\mathbf{\mathsf \Phi} = \mathbf{\mathsf M}_{0.12}\mathbf{\mathsf F}$ & (c)  $\mathbf{\mathsf \Phi} = \mathbf{\mathsf M}_{0.09}\mathbf{\mathsf F}$       \\
			\includegraphics[trim={{.2\Lwidth} {.17\Lwidth} {.04\Lwidth} {.12\Lwidth}}, clip, width=\ww, height=\hhhh]{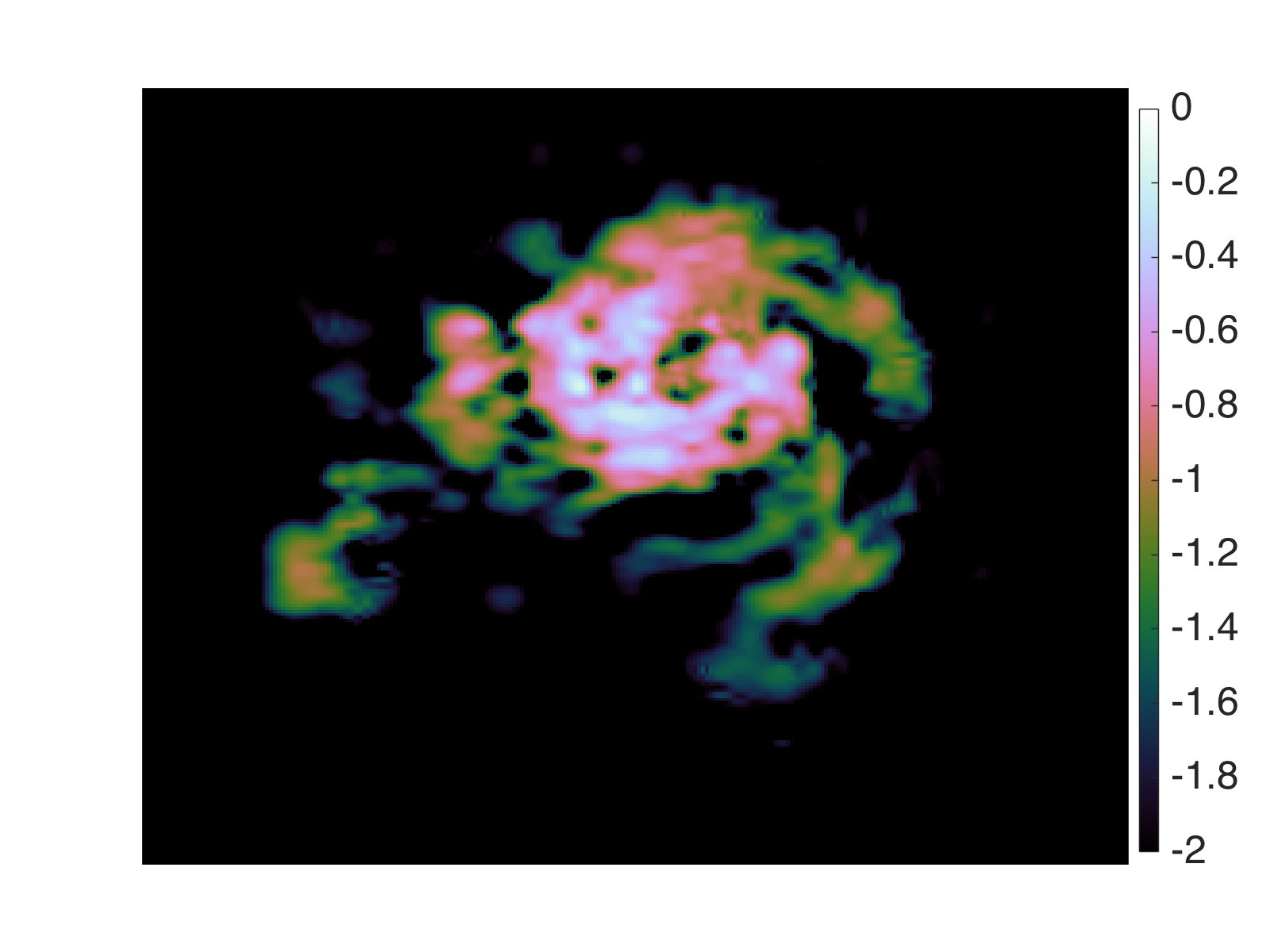} &
			\includegraphics[trim={{.2\Lwidth} {.17\Lwidth} {.04\Lwidth} {.12\Lwidth}}, clip, width=\ww, height=\hhhh]{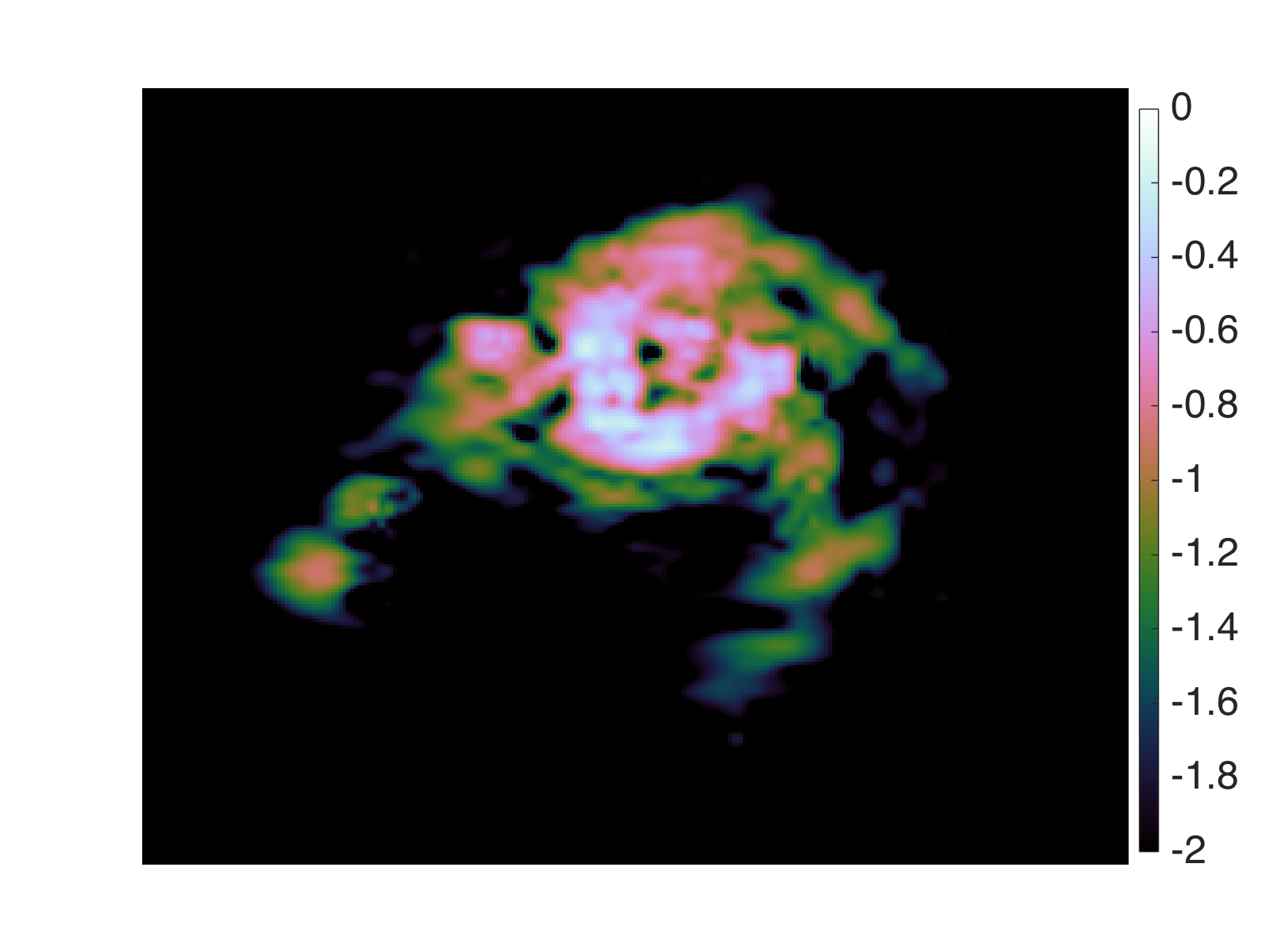} &
			\includegraphics[trim={{.2\Lwidth} {.17\Lwidth} {.04\Lwidth} {.12\Lwidth}}, clip, width=\ww, height=\hhhh]{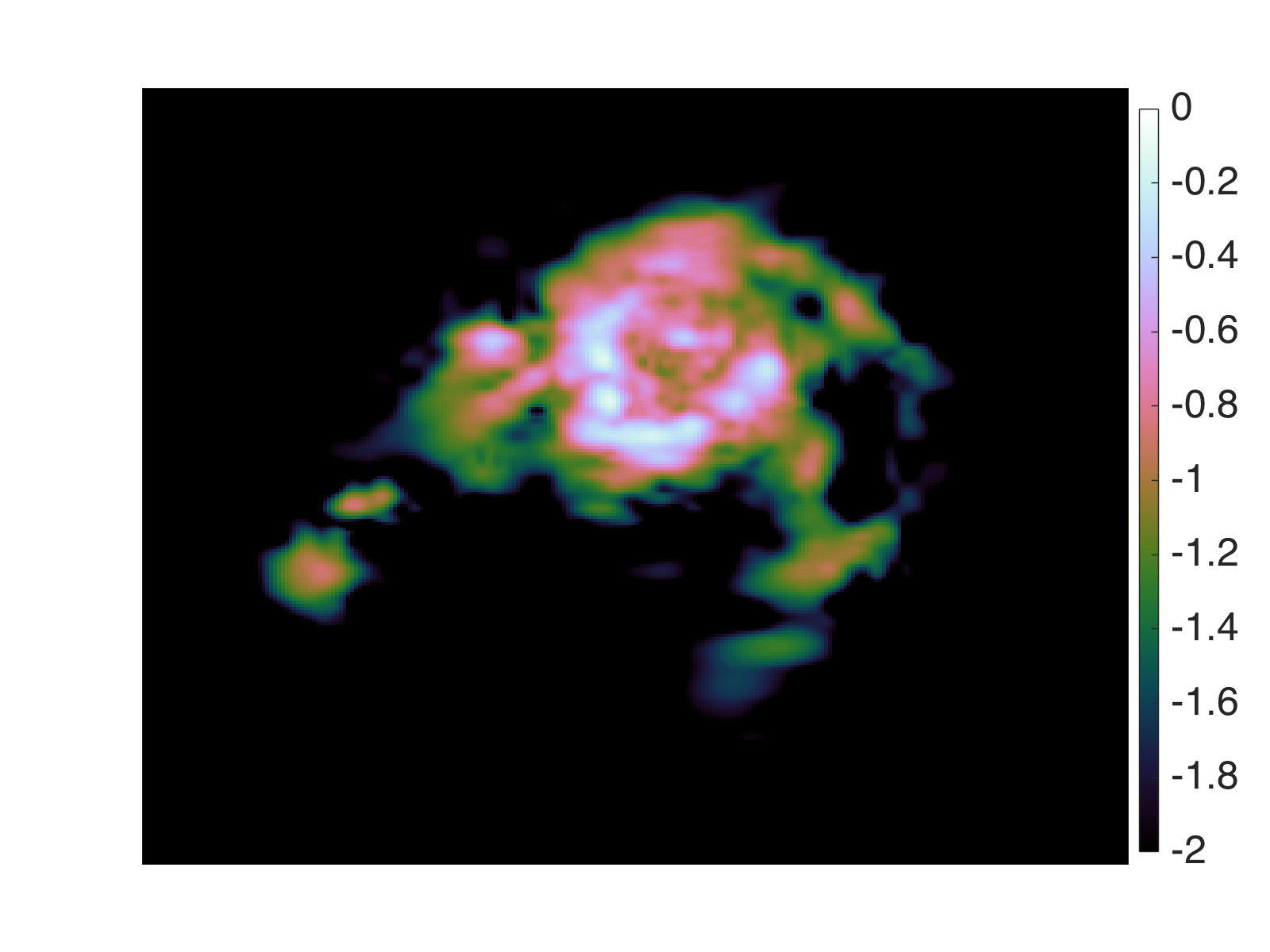}                                                                                                                                                                   \\
			(d)  $\mathbf{\mathsf \Phi} = \mathbf{\mathsf M}_{0.06}\mathbf{\mathsf F}$                                                                                                                                                         & (e) $\mathbf{\mathsf \Phi} = \mathbf{\mathsf M}_{0.03}\mathbf{\mathsf F}$ & (f)  $\mathbf{\mathsf \Phi} = \mathbf{\mathsf M}_\text{truth}\mathbf{\mathsf F}$
		\end{tabular}
	\end{center}
	\caption{Measurement model misspecification for an image reconstruction problem solved by proximal nested sampling (test image is M31 radio galaxy).
		Panel (a): dirty (back-projected) image $\mathbf{\mathsf \Phi}^\dagger Y$; Panels (b)--(f): posterior mean images recovered by proximal nested sampling for misspecified models $\mathbf{\mathsf M}_\gamma$, where increasing $\gamma > 0$ corresponds to increasing levels of misspecification (and $\gamma=0$ corresponds to the ground truth model).  It is apparent by eye that the posterior mean image recovered with the ground truth model is the best and that the quality of the recovered posterior mean image degrades as the size of the misspecification scale parameter $\gamma$ increases.
	}\label{Fig:test-ms-recons-M31}
\end{figure*}
%% ---------------------------------------

Fig.~\ref{Fig:test-ms-recons-M31} presents the posterior means recovered (i.e.\
the reconstruction images)
for models with $\mathbf{\mathsf \Phi}_\gamma = \mathbf{\mathsf{M}}_\gamma \mathbf{\mathsf{F}}$ and $\mathbf{\mathsf \Phi} = \mathbf{\mathsf{M}}_\text{truth}\mathbf{\mathsf{F}}$,.
Here misspecified models $\mathbf{\mathsf M}_{0.12}$, $\mathbf{\mathsf M}_{0.09}$, $\mathbf{\mathsf M}_{0.06}$ and $\mathbf{\mathsf M}_{0.03}$ are generated
for misspecification scaling factors $\gamma$ with values of 0.12, 0.09, 0.06 and 0.03, respectively.
It is apparent by eye that the posterior mean image recovered with the ground truth model is the best and that the quality of the recovered posterior mean image degrades as the size of the misspecification scale parameter $\gamma$ increases.
The computation time for each problem is approximately $150$ minutes.

%% ---------------------------------------
\begin{table}[!h]
	\centering \caption{Marginal likelihood (Bayesian evidence values computed by proximal nested sampling for Bayesian model selection for measurement model misspecification for an image reconstruction problem (see Fig.~\ref{Fig:test-ms-recons-M31} for corresponding reconstructed images).  Misspecified models are denoted $\mathbf{\mathsf M}_\gamma$, where increasing $\gamma > 0$ corresponds to increasing levels of misspecification (and $\gamma=0$ corresponds to the ground truth model).  Comparing models, Bayesian model selection afforded by proximal nested sampling suggests the model with the lowest misspecification parameter $\gamma$ is always preferred, which also agrees with the RMSE (root mean square error) values and assessment performed by eye, which require the ground truth to be known.
	}
	\begin{tabular}{lcr}
		\toprule
		\multicolumn{1}{c}{Likelihood}                                              & $\log {\cal Z}$              & RMSE  \\ \midrule
		$\mathbf{\mathsf \Phi} = \mathbf{\mathsf M}_\text{truth}\mathbf{\mathsf F}$ & $-4.47\times 10^3{\pm0.08} $ & 3.40  \\
		$\mathbf{\mathsf \Phi} = \mathbf{\mathsf M}_{0.03}\mathbf{\mathsf F}$       & $-4.88\times 10^3{\pm0.08}$  & 7.85  \\
		$\mathbf{\mathsf \Phi} =  \mathbf{\mathsf M}_{0.06}\mathbf{\mathsf F}$      & $-5.63\times 10^3{\pm0.08} $ & 12.01 \\
		$\mathbf{\mathsf \Phi} =  \mathbf{\mathsf M}_{0.09}\mathbf{\mathsf F}$      & $-9.21\times 10^3{\pm0.07} $ & 15.71 \\
		$\mathbf{\mathsf \Phi} =  \mathbf{\mathsf M}_{0.12}\mathbf{\mathsf F}$      & $-1.44\times 10^4{\pm0.08} $ & 18.08 \\\bottomrule
	\end{tabular}
	\label{table-BE-recons-M31}
\end{table}
%% ---------------------------------------

Table~\ref{table-BE-recons-M31} presents the marginal likelihood and RMSE values computed for the different models considered.
The computed marginal likelihood is largest when the correct ground truth model is adopted in the likelihood.  As the misspecification parameter $\gamma$ is increased (corresponding to greater misspecification and less accurate models), the corresponding computed marginal likelihood values monotonically decrease (become more negative).  For Bayesian model comparison, the model with the lowest misspecification parameter $\gamma$ is always preferred.
The computed marginal likelihoods are consistent with the model preferences obtained by comparing the RMSE of each model and by visual inspection (although recall that such tests cannot be used for model comparison in practice when the ground truth is not known).

%%---------------------
%\subsection{Further discussion} %\label{sec:further-diss}
%%---------------------
%\red{Before closing this section, we further discuss %some other existing methods and their
%relationship and difference with the proposed %proximal nested sampling. }

%\red{
%Nested sampling with constrained Hamiltonian Monte Carlo was proposed in \cite{B11}...
%Gradient importance sampling was proposed in \cite{S15}...
%Gibbs sampling scheme with non-parametric methods \xc{Not sure which recent papers for this type of methods we
%should cite.}
%}

%-------------------------------------------------------------------
\section{Conclusions}\label{sec:conclusions}
%-------------------------------------------------------------------
Nested sampling provides an efficient computational framework to estimate the marginal likelihood (Bayesian evidence) for Bayesian model selection.  It effectively re-parameterises the marginal likelihood into a one-dimensional integral of the likelihood with respect to the enclosed prior volume.  The challenge of nested sampling is to sample from the prior distribution subject to a hard likelihood constraint.  A variety of successful techniques have been developed to perform such sampling in low and moderate dimensional problems. However, existing approaches are not directly useful for imaging applications because they scale poorly to large problems and struggle to support models that are not smooth.

In this article we presented the proximal nested sampling method that is specifically designed for Bayesian models that are log-concave, potentially very high-dimensional ($d = 10^6$ and beyond), and potentially not smooth. This is achieved by exploiting tools from proximal calculus and Moreau-Yosida regularisation to efficiently sample from the prior subject to the hard likelihood constraint through a proximal MCMC approach. The resulting Markov chain iterations combine a gradient step that approximates a Langevin SDE that scales efficiently to large problems, with a projection term that acts to push the Markov chain back into the likelihood constraint set if it wanders outside of it, and a Metropolis-Hastings correction step to ensure the hard likelihood constraint is satisfied.

The proposed proximal nested sampling framework was implemented and validated on a Gaussian model for which the marginal likelihood could be calculated in closed-form, showing excellent agreement between values computed analytical and by proximal nested sampling, even in very high dimensions. The use of proximal nested sampling for principled Bayesian model selection was then showcased on a variety of imaging problems with non-smooth sparsity-promoting prior distributions. In particular, model selection problems were considered related to dictionary selection, and selection of the appropriate measurement model when it may be misspecified.

Proximal nested sampling allows Bayesian model selection to be performed at a much higher dimension than that was previously possible, while also supporting non-smooth priors that are widely used in imaging. It is our hope that proximal nested sampling will thus find widespread use for high-dimensional Bayesian model selection, particularly in the imaging sciences.

{Important perspectives for future work include: a detailed theoretical analysis of the convergence properties of proximal nested sampling; an extension to (biased) accelerated proximal methods \citep{vargas2019accelerating}; and an analysis of the properties of marginal maximum likelihood estimation for the class of models considered in this paper, such as estimator consistency for model selection in an $\mathcal{M}$-closed setting and concentration in an $\mathcal{M}$-open setting \citep{LMCLD22}. Moreover, it would be interesting to apply proximal nested sampling to other types of models, such as models with likelihood-based priors \citep{LMCLD22}, which can be handled straightforwardly by proximal nested sampling when the likelihood is log-concave. It would also be interesting to modify proximal nested sampling to tackle high-dimensional models that are multi-modal, particularly models with data-driven priors encoded by neural networks (see e.g. \citealt[Section 5]{Mukherjee2022})}.

%%%%%%%%%%%%%%%%%%
%% use section* for acknowledgment
%\section*{Acknowledgment}
% This work was supported by the Leverhulme Trust and by EPSRC (EP/T007346/1).
% %
% For the purpose of open access, the authors have applied a Creative Commons Attribution (CC BY) licence to any Author Accepted Manuscript version arising.

%%%%%%%%%%%%%%%%%%%%%%%%%
\begin{appendices}

	\section{}\label{secA1}

	The volume of the prior $f(x) = \mu \|\mathbf{\mathsf \Psi}^\dagger x\|_2^2$ with $\mathbf{\mathsf \Psi} = I$ is
	\begin{align}
		\begin{split}
			V = & \int_{-\infty}^{\infty} \exp\left(-\mu \|x\|_2^2 \right) \text{d} x  \\
			= & 	\int_{-\infty}^{\infty} \exp\left(-\frac{1}{2} 2\mu  x^\top x \right) \text{d} x  \\
			= &  \sqrt{\frac{(2\pi)^d}{(2\mu)^d}}.
		\end{split}
	\end{align}

	For the prior $f(x) = \mu \|\mathbf{\mathsf \Psi}^\dagger x\|_2^2$ with $\mathbf{\mathsf \Psi} = I$ and the likelihood $g(x) = \|y -\mathbf{\mathsf \Phi} x\|_2^2/{2\sigma^2}$ with
	$\mathbf{\mathsf \Phi} = I$, the Bayesian evidence value has the following closed-form representation:
	\begin{align}
		\begin{split}
			& \frac{1}{V} \int_{-\infty}^{\infty} \exp(-\mu \|x\|_2^2) \exp(-\|y - x\|_2^2/{2\sigma^2}) \text{d} x \\
			= & \frac{1}{V} \int_{-\infty}^{\infty} \exp\left(-\mu \|x\|_2^2 - \|y - x\|_2^2/{2\sigma^2}\right) \text{d} x  \\
			= & \frac{1}{V} \int_{-\infty}^{\infty} \exp\left(-(\mu + 1/2\sigma^2) x^\top x + y^\top x/\sigma^2 - y^\top y/2\sigma^2\right) \text{d} x  \\
			= & \frac{1}{V} \exp\left(- \frac{y^\top y}{2\sigma^2}\right)
			\int_{-\infty}^{\infty} \exp\left(-\frac{1}{2}(2\mu + 1/\sigma^2) x^\top x + y^\top x/\sigma^2\right) \text{d} x  \\
			= & \frac{1}{V} \sqrt{\frac{(2\pi)^d}{(2\mu + 1/\sigma^2)^d}} \exp\left(- \frac{y^\top y}{2\sigma^2}\right)
			\exp\left(\frac{1}{2} \frac{1}{2\mu + 1/\sigma^2} \frac{y^\top y}{\sigma^4}  \right),
		\end{split}
	\end{align}
	whose logarithmic value is
	\begin{align}
		\begin{split}
			\log\sqrt{\frac{(2\pi)^d}{(2\mu + 1/\sigma^2)^d}} + \left(- \frac{y^\top y}{2\sigma^2}\right)
			+ \left(\frac{1}{2} \frac{1}{2\mu + 1/\sigma^2} \frac{y^\top y}{\sigma^4}  \right) - \log V.
		\end{split}
	\end{align}




\end{appendices}


\backmatter

\bmhead{Acknowledgments}
This work was supported by the Leverhulme Trust and by EPSRC grants EP/T007346/1 and EP/W007673/1. The authors would like to thank the editor and two anonymous reviewers for their valuable suggestions to improve the manuscript. The authors are also grateful to Abdul-Lateef Haji-Ali for helpful comments.
For the purpose of open access, the authors have applied a Creative Commons Attribution (CC BY) licence to any Author Accepted Manuscript version arising.

\bibliography{refs_xhcai}% common bib file


\end{document}